\documentclass[lettersize,journal]{IEEEtran}
\usepackage{amsmath,amssymb,booktabs,graphicx,xcolor,url}
\graphicspath{{./}}
\usepackage{cite}

\newcommand{\kd}{\kappa}
\newtheorem{proposition}{Proposition}
\newtheorem{corollary}{Corollary}
\newtheorem{conjecture}{Conjecture}

\begin{document}

\title{The Unreported Variable:\\Collector Selectivity in Data Placement Evaluation}

\author{Nannan Zhao, Jiameng Zhang, Taoyu Zhong, Bo Wang, Hao Wang, Zhijie Huang, Shujie Han, Xiaonan Zhao, Xiao Zhang%
\thanks{Nannan Zhao, Jiameng Zhang, Taoyu Zhong, Bo Wang, Hao Wang, Zhijie Huang, Shujie Han, Xiaonan Zhao and Xiao Zhang are with NorthWestern Polytechnical University.}%
\thanks{Corresponding author: Nannan Zhao(e-mail: nannanzhao@nwpu.edu.cn).}%
}

\maketitle

\begin{abstract}
A solid-state drive cannot overwrite data in place, so it cleans full blocks by copying out still-valid data, costing speed and lifetime. Interfaces such as NVMe Flexible Data Placement (FDP) let the host label data by expected lifetime so data dying together is stored together, and published schemes report cutting these copies by 20--75\%. Those studies rarely state which blocks the drive chooses to clean. Varying only this choice in simulation, we show it decides how much a label is worth: label value peaks when the drive searches moderately for blocks holding little valid data, falls when it always takes the block with the least, and turns negative when it ignores how much valid data blocks hold. This holds in 47 of 48 synthetic configurations, and one real mail-server trace shows the same shape at about half the magnitude. Priced as spare capacity, a two-class label is worth 65\% more spare when each scheme runs at its best setting, but becomes a net loss at 20\% label error, typical of learned classifiers. Labels stating which data die together keep most of their value at 16 FDP handles and gain as cleaning improves. Evaluations must report the drive's cleaning policy.
\end{abstract}

\begin{IEEEkeywords}
Data placement, Flash memory, Garbage collection, Hot/cold data separation,
NVMe Flexible Data Placement, Over-provisioning, Victim selection, Write amplification.
\end{IEEEkeywords}

%
%
\section{Introduction}
\label{sec:intro}

Flash memory is programmed in pages but erased in blocks of many pages, so a
drive cannot update data in place. When free space runs low, a garbage collector
picks a fully written block --- the victim --- copies out the pages that are
still valid, and erases it. Those copies are write amplification, and they cost
both throughput and endurance~\cite{agrawal2008tradeoffs,hu2009writeamp}.
How much the collector must copy depends on two choices: \emph{which data are
stored together}, and \emph{which block the collector reclaims}.

The first choice has had a decade of attention. Multi-streamed SSDs, Zoned
Namespaces (ZNS) and NVMe Flexible Data Placement (FDP) let the host influence
which data share an erase unit~\cite{kang2014multistream,bjorling2021zns,nvme_fdp},
and a long line of work supplies the labels that steer it --- drawn from access
frequency, recency, update counts, migration history, program context, file
type, or learned predictors --- and reports reductions of 20--75\%
(Table~\ref{tab:roster}). With FDP now part of the NVMe specification and
offered by commercial drives~\cite{nvme_fdp,warp2026}, hosts are being asked to
supply such labels, and whether that effort pays is judged from these reported
gains.

The second choice has been studied for its own sake but not for its effect on
labels. Analytical work shows that the victim-selection policy changes the
write amplification of each placement scheme, and that greedy selection is not
optimal once hot and cold data are separated~\cite{vanhoudt2013hotcold}. What
has not been measured is whether the policy changes the \emph{difference}
between schemes --- that is, how much a label is worth. Placement evaluations
rarely report the policy they ran under, and under FDP it belongs to the device,
which does not expose it. Two recent observations suggest this matters.
Cross-device characterisation of commercial FDP drives finds the benefit real
but unstable, with the same workload differing by nearly $2\times$ between
devices~\cite{warp2026}; and the most recent placement system compares arms that
run under three different victim-selection policies, so its reported gain
combines a change of placement with a change of collector~\cite{dogi2026}.
Neither treats the collector as a variable, and that practice is the subject of
this paper.

\noindent\textbf{The question.} How much of what a label is worth is decided by
the collector? We hold the workload, the labels, the device and the spare
capacity fixed and vary only victim selection: from uniform-random choice,
through $d$-choices, which samples $d$ candidate blocks and takes the one with
fewest valid pages, to greedy, which takes the fewest over all candidates, and
finally FIFO, which reclaims in write order and ignores occupancy altogether.
Three write schemes span the separation a device can
perform~\cite{vanhoudt2014necessity}: none (SWF); host writes kept apart from
relocated data, which separates young from old with no label at all (DWF); and a
two-class label (HCWF). Section~\ref{sec:method} defines all seven precisely.
Table~\ref{tab:accounts} summarises the outcome on one workload;
Section~\ref{sec:collector} gives the full sweeps behind it.

\begin{table}[htbp]
\caption{Write amplification of each scheme under each victim-selection policy
($r{=}0.9$, $f{=}0.05$, $S_f{=}0.10$). ``Best $d$'' is the sampling width that
minimises that scheme's own write amplification (\S\ref{sec:method}).}
\label{tab:accounts}
\centering
\footnotesize
\begin{tabular}{lrrrr}
\toprule
& \multicolumn{4}{c}{victim-selection policy} \\
\cmidrule(l){2-5}
scheme & uniform & greedy & best $d$ & FIFO \\
\midrule
SWF~\cite{hu2009writeamp,bux2010greedy,desnoyers2014analytic,vanhoudt2013meanfield}  & 9.950 & 5.820 & 5.787 & 6.93 \\
DWF~\cite{vanhoudt2013hotcold}  & 10.031 & 4.721 & 3.004 & 6.98 \\
HCWF~\cite{vanhoudt2014necessity} & 10.042 & 3.866 & 2.093 & $17.3\pm1.5$ \\
\bottomrule
\end{tabular}
\end{table}

\noindent\textbf{Three accounts, not one.}
Read with each scheme under the collector that suits it best,
Table~\ref{tab:accounts} splits placement benefit into three terms.
\emph{Host separation}, DWF over SWF and needing no label, is worth
$1.93\times$. \emph{The label}, HCWF over DWF, is worth $1.44\times$.
\emph{The collector} --- the same scheme, DWF, under a policy that samples a
handful of candidates rather than searching them all, which is the greedy
default the literature assumes --- is worth $1.57\times$, and $2.10\times$ on
our second workload. The terms are not commensurable: the first two compare
schemes under one policy, the third compares policies for one scheme, so we do
not rank them. The point is that the third is of the same order as the other
two, is a property of neither the workload nor the labels, and, to our
knowledge, is reported by no published placement evaluation; under FDP the host
cannot even observe it.

\noindent\textbf{What we find, and why.} The value of a label is not a property
of the label. Measured as the share of eliminable write amplification it
removes, $g$ (\S\ref{sec:metric}), a two-class label peaks at $0.506$ near
$d{=}24$, falls to $0.230$ under greedy, and vanishes under uniform selection;
under FIFO it turns negative and label-free schemes win, so which scheme one
should deploy depends on the collector. The shape holds in 47 of 48 simulated
configurations, and on a real mail-server trace it reproduces across three seeds
and two successive 50M-write windows (\S\ref{sec:collector}). The reason is what
the collector can see: a collector that only chooses victims observes each
block's valid count and age, and nothing about which pages those are
(\S\ref{sec:theory}). Rate labels therefore restate what a searching collector
already infers --- once the partition is correct they add at most about 1\% ---
while membership labels, which say which pages die together, carry what it
cannot reconstruct, and keep most of their value at the sixteen handles a
commercial drive exposes. Priced in the currency an operator buys, extra spare
capacity, a two-class label is worth about two-thirds more capacity when each
scheme runs at its own best setting, but turns negative at the error rates
learned classifiers report (\S\ref{sec:labels}).

\noindent\textbf{What this means for prior work.} Prior results are incomplete
rather than wrong: a placement gain reported without the collector cannot be
carried to another device, and under FDP the host can neither observe the
collector nor select the label-free baseline against which its labels should be
judged.

In this paper, we make the following contributions:
\begin{itemize}
\item \textbf{A controlled measurement}, to our knowledge the first, of how
victim selection changes the value of placement labels, with workload, labels,
device and spare capacity held fixed (\S\ref{sec:collector}).
\item \textbf{A decomposition} of placement benefit (Table~\ref{tab:accounts})
that identifies collector selectivity as an unreported term of the same order as
host separation and labels.
\item \textbf{A theory} of when a label adds information: permutation
invariance, a conditional-information bound, a class-count condition, and a
scale ratio (\S\ref{sec:theory}).
\item \textbf{A price} in spare capacity, under realistic label error and at
deployable handle counts, separating the labels worth supplying from those that
are not (\S\ref{sec:labels}), and the evaluation practice that follows
(\S\ref{sec:deliver}).
\end{itemize}

\noindent The rest of the paper is organised as follows.
Section~\ref{sec:related} reviews garbage collection and placement.
Section~\ref{sec:framework} sets out the device model, the metric, the theory and
the workloads. Section~\ref{sec:collector} measures how label value depends on
the collector, and Section~\ref{sec:labels} what labels are worth and which kinds
pay. Section~\ref{sec:deliver} concludes and states the limits. Sections
numbered S$n$ are in the supplemental material, which also holds the
pre-registered verdict rules and the corrections made to earlier drafts.

%
%
\section{Background and Related Work}
\label{sec:related}

The problem has two halves with separate literatures. Work on the collector
models which blocks to reclaim, usually without labels; work on placement
designs labels, usually without stating the collector. This section reviews
each in turn, and shows where they fail to meet.

\subsection{Garbage collection and victim selection}

Updates are made out of place, so before erasing a block the collector must
relocate whatever is still valid in it. The ratio of media writes to host
writes, the write amplification, is a first-order determinant of both throughput
and lifetime, since blocks tolerate a bounded number of program/erase
cycles~\cite{agrawal2008tradeoffs,hu2009writeamp}. We assume page-level mapping
throughout, as the analytical literature does.

The choice of victim has a long literature of its own, and it is the variable
this paper sweeps. \textbf{FIFO} reclaims blocks cyclically and is blind to
occupancy~\cite{rosenblum1992lfs}. \textbf{Greedy} reclaims the block with fewest
valid pages and is believed optimal under uniform random
writes~\cite{bux2010greedy}. \textbf{Cost-benefit} weighs the space freed
against the copy cost and the block's age, using age as a proxy for future
invalidation~\cite{rosenblum1992lfs}. \textbf{Windowed} greedy restricts the
search to the $w$ least recently selected blocks~\cite{bux2010greedy}.
\textbf{$d$-choices} samples $d$ blocks at random and reclaims the one with
fewest valid pages, and
around ten samples already approach greedy under uniform
writes~\cite{vanhoudt2013meanfield}.

These policies are usually presented as a trade-off between complexity and
quality, with greedy as the ceiling and the rest as cheaper approximations. Two
results complicate that picture. First, once hot and cold data are present and
a double write frontier is used, greedy is \emph{not} optimal: an intermediate
$d$ is better, because a block holding mostly hot data gains from being left to
self-invalidate~\cite{vanhoudt2013hotcold}. Second, under uniform writes greedy
is also near-optimal for endurance without extra wear
levelling~\cite{verschoren2016endurance}. Both are observations about the
collector, not about any host policy, and both hint that the collector does more
of the work than placement studies credit.

For uniform random writes the analysis is essentially closed. A Markov-chain
formulation gives exact write amplification for greedy on small and medium
devices, with a large-system analysis in which the valid count of a reclaimed
block concentrates on two consecutive values~\cite{bux2010greedy}. Closed forms
follow for LRU cleaning, via the Lambert $W$ function, with a greedy
approximation derived from it~\cite{desnoyers2014analytic} and later
refined~\cite{xiang2012improved}. Mean-field analysis covers a class of
collectors including $d$-choices~\cite{vanhoudt2013meanfield} and extends to
endurance~\cite{verschoren2016endurance}. A companion analysis treats $K$
lifetime classes, derives the optimal division of spare capacity in closed form,
and bounds the placement ceiling by a Bhattacharyya coefficient, validated to
0.3\% under random victim selection~\cite{companion}. Every model in this family carries only \emph{first moments} --- rates per class
and counts per block. Section~\ref{sec:theory} argues that this is not a
limitation of the models but a correct reflection of what a victim-selecting
collector can act on. We claim no novelty in modelling collection as a Markov or
mean-field process~\cite{vanhoudt2013meanfield,benaim2008meanfield}; what is
absent there is a two-time-scale separation --- fast page churn within a block,
slow filling and emptying of blocks --- which is the limit in which
Proposition~\ref{prop:perm} holds.

\subsection{Placement interfaces and lifetime labels}

Multi-streamed SSDs let the host tag writes with a stream identifier so that
data of different expected lifetimes land in different
blocks~\cite{kang2014multistream}. ZNS exposes zones as the placement unit,
requires sequential writes, and moves collection to the
host~\cite{bjorling2021zns}. FDP generalises the idea through reclaim unit
handles while keeping the conventional block interface~\cite{nvme_fdp}. The
distinction that matters here is \emph{ownership}. Under ZNS the host owns
placement \emph{and} collection, as in ZenFS for RocksDB and in zoned
F2FS~\cite{bjorling2021zns,lee2015f2fs}. Under FDP the host owns only the label;
the device chooses victims under a vendor-defined policy that it does not expose.
Characterisation of commercial devices reports that ZNS zone management carries
substantial and uneven cost~\cite{doekemeijer2023zns}, and that FDP benefit is
unstable across workloads and drives~\cite{warp2026}.

\begin{table}[t]
\caption{Lifetime-identification schemes and the baselines they report against.}
\label{tab:roster}
\centering
\small
\setlength{\tabcolsep}{3pt}
\begin{tabular}{llcc}
\toprule
scheme & signal & gain & baseline \\
\midrule
hash counters~\cite{hsieh2006hotdata}          & frequency        & ---      & --- \\
Bloom filters~\cite{park2011bloom}             & recency          & ---      & --- \\
dynamic clustering~\cite{chiang1999dac}        & update freq.\     & ---      & --- \\
update frequency~\cite{stoica2013update}       & write skew       & 20--75\% & single \\
extent-based~\cite{shafaei2016eti}             & extent heat      & 20\%     & single \\
migration counts~\cite{park2021mida}           & relocations      & 67\%     & single \\
automatic streams~\cite{yang2017autostream}    & freq.\ + recency & 60\%     & single \\
program contexts~\cite{kim2019pcstream}        & call stack       & ---      & --- \\
oracle-guided~\cite{dogi2026}                  & learned          & 56--84\% & acc.\ only \\
\bottomrule
\end{tabular}
\end{table}

Table~\ref{tab:roster} collects the identification line, from counters and
filters in the host to classification inside firmware and stream assignment.
Each reports a gain on the workload it was evaluated on, except the last, which
reports classification accuracy. Where a baseline is stated it
is a single write frontier (``single''), which performs no separation at all.
Since a double write frontier performs the coarsest hot/cold split with no
labels~\cite{vanhoudt2014necessity}, these figures include a split that is free.
The most recent systems do perform some such separation, so their gains are not
purely the free split relabelled; what is missing is a \emph{stated, controlled}
reference that performs the free split and nothing else.

\subsection{Invalidation-time placement, and a current instance}
\label{sec:dogi}

A separate line places data by expected invalidation time rather than access
rate. An optimal placement exists given each block's future invalidation time,
and can be approximated online from write skew~\cite{wang2022sepbit}; a recent
system uses a near-optimal oracle as an upper bound and measures a learned
predictor against it~\cite{dogi2026}. This view treats the collector as fixed.
It complements ours: the invalidation-time optimum is the \emph{ceiling}, a
double write frontier is the \emph{free part}, and our subject is the residual
between them and how it moves with the collector.

That system also illustrates the reporting gap, and we describe it from the
published text without re-running it. Its evaluation uses three different
victim-selection policies across the arms it compares: several baselines use
cost-benefit, the principal baseline combines FIFO with cost-benefit, and the
proposed scheme uses its own rule, reclaiming an expired segment if one exists
and otherwise the segment with fewest valid blocks from a designated
group~\cite{dogi2026}. It
reports a 15.5\% average reduction, up to 23.2\%, over that baseline. Two
variables therefore differ between the arms, and Section~\ref{sec:collector}
shows that the second can move write amplification as much as the first. The
same work finds an interior optimum in the number of groups, which moves with
the prediction method~\cite{dogi2026}; that is an instance of the first-order
condition of Proposition~\ref{prop:foc}, so a sweet spot in class count is not
ours to claim. If its gain survives at matched victim selection, as we would
expect, the attribution stands and our point narrows to reporting practice
(\S S5).

Our platform builds on the flash-emulation line~\cite{li2018femu} and public
block traces~\cite{narayanan2008writeoffload}. Flash caching supplies the
clearest published demonstrations of placement benefit, including device-level
amplification near unity with FDP~\cite{allison2025fdpcache}.

%
%
\section{Measuring Label Value: Model, Metric, and Theory}
\label{sec:framework}

To separate what the collector contributes from what a label contributes, we
need four things, and this section supplies them in order. The first is a device
model in which the collector and the labels can be varied independently
(\S\ref{sec:method}). The second is a measure of label value whose scale does
not itself change with the collector (\S\ref{sec:metric}). The third is a
prediction of how that value should depend on what the collector can see
(\S\ref{sec:theory}--\S\ref{sec:regime}). The last is a set of workloads and
checks that let the measurements of Sections~\ref{sec:collector}
and~\ref{sec:labels} be trusted (\S\ref{sec:workloads}).

\subsection{Device model, write schemes and collectors}
\label{sec:method}

We simulate a page-mapped device without TRIM: $N$ blocks of $b$ pages, of which
a fraction $S_f$ (the spare factor) is not addressable by the host. Unless
stated otherwise $N{=}600$, $b{=}32$ and $S_f{=}0.10$. A run writes the logical
capacity a stated number of times (passes), and write amplification is measured
after the run reaches steady state. The simulator asserts its own bookkeeping
throughout: physical writes equal host plus relocation writes, every logical page
has one valid version, and per-block valid counts are non-negative and sum to
the number of mapped pages.

The three write schemes are those of the 2014 study~\cite{vanhoudt2014necessity}.
SWF has one open block (a \emph{write frontier}) and writes the survivors of a
reclaimed block back into the same stream as host data. DWF keeps two frontiers,
one for host writes and one for relocated data. HCWF keeps a hot and a cold
frontier: host writes go by their label, survivors go to the frontier matching
the victim's label, and a block inherits the label of the frontier it last
served. For class-count experiments we extend DWF to $\mathrm{KWF}(K)$, with $K$
labelled data frontiers plus one frontier for relocated data, so that
$\mathrm{KWF}(1)$ is exactly DWF.

\begin{table}[t]
\caption{The collectors we sweep. Every candidate is a fully written block, so
the policies differ only in how much valid data the block they pick still
holds.}
\label{tab:collectors}
\centering
\footnotesize
\setlength{\tabcolsep}{4pt}
\begin{tabular}{lll}
\toprule
policy & what it looks at & victim \\
\midrule
uniform ($d{=}1$) & nothing            & a candidate at random \\
$d$-choices       & $d$ sampled counts & fewest valid of the $d$ \\
greedy            & all counts         & fewest valid overall \\
FIFO              & allocation order   & oldest written \\
\bottomrule
\end{tabular}
\end{table}

The collector is the variable, and Table~\ref{tab:collectors} gives the four we
sweep; larger $d$ is more selective. In our device there are 599 candidates for
SWF and 598 for DWF and HCWF once open frontiers are excluded, and we verify
that $d$ equal to the candidate count takes the greedy code path and gives
identical write amplification. FIFO admits choices that change its behaviour,
and we fix them: one global queue over all blocks, and a block whose survivors
are written back does not move in the queue. A per-frontier queue would cycle a
cold frontier at its own fill rate and should worsen the effect of
\S\ref{sec:fifo}; we have not swept it.

Absolute $d$ does not transfer between devices: a drive with $N \sim 10^{5}$
blocks samples a far smaller fraction at the same $d$. A natural transferable
form is the probability of sampling at least one structure-carrying block,
$1-(1-\phi)^d$, which predicts $d^\ast \sim 1/\phi$. Separating the workload
parameters refutes it: with the hot-page fraction held fixed, $d^\ast$ still
moves by a factor of two as the traffic share varies. We therefore report $N$
and the workload with every configuration, and our claims concern the existence
and size of the dependence on $d$, not where its optimum lies on another drive.

Class-count and statistic sweeps use a second discrete-event simulator with the
same assertions. Every derived number --- $g$, net value, $\Delta S/S$, $K^\ast$
--- is computed by one script from a stated configuration, which also asserts
that $\mathrm{KWF}(1)$ reproduces DWF and suppresses $g$ outside its domain
(\S\ref{sec:metric}). This guards against the kind of error that three of our
earlier corrections fixed: derived quantities assembled by hand from runs made
under different configurations (\S S5). All verdict rules were fixed before the corresponding runs and are listed with
their outcomes in \S S4; the artifact ships the scripts that produced each
table.

\subsection{Measuring label value}
\label{sec:metric}
\label{sec:comparison}

Placement benefit is conventionally reported as a ratio,
$1 - A_{\text{policy}}/A_{\text{base}}$, where $A$ is write amplification. That
ratio has a ceiling that changes with the configuration. No policy can drive
amplification below one, so the numerator cannot exceed $A_{\text{base}}-1$, and
\begin{equation}
1-\frac{A_{\text{policy}}}{A_{\text{base}}}
=\underbrace{\frac{A_{\text{base}}-A_{\text{policy}}}{A_{\text{base}}-1}}_{\textstyle g}
\times\underbrace{\Bigl(1-\frac{1}{A_{\text{base}}}\Bigr)}_{\text{ceiling}} .
\label{eq:g}
\end{equation}
Across the configurations of Tables~\ref{tab:collector} and~\ref{tab:g} the
ceiling runs from $0.337$ to $0.900$, a factor of $2.7$. Ranking configurations
on a scale whose maximum moves that much must mis-rank them, and always in the
same direction: it favours configurations with poor absolute amplification, that
is, weak collectors. We therefore report $g$, the share of \emph{eliminable}
amplification a policy removes. This is a standard normalisation, not a new
index. Unless stated otherwise the policy is HCWF, the baseline is DWF, and $g$
measures what the label adds beyond the free split.

$g$ has a domain. As $A_{\text{base}}\to1$ its numerator and denominator both
vanish. A first-order bound on its variance, proportional to
$(A_{\text{base}}-1)^{-2}$, is far too pessimistic: it predicts the standard
deviation of $g$ growing $9.4\times$ between $S_f{=}0.05$ and $0.30$, where
five seeds give $2.1\times$, non-monotonically (\S S3). We therefore report
measured dispersion, report $g$ only where $A_{\text{base}}\ge1.2$, and give
$A_{\text{base}}$ wherever a $g$ supports a claim. Outside that domain $g$ means
nothing: at a run length too short for steady state, four cells of the robustness
grid had $A_{\text{base}}\approx1.16$ and an apparent $g\approx0.91$.

An operator asks a different question: is a label a better purchase than more
flash? We answer it with a \emph{capacity equivalent}. For a labelled scheme
running at $S_f{=}0.10$ we find the spare factor $S^\ast$ at which the
label-free DWF reaches the same write amplification, and report
$\Delta S/S = (S^\ast-0.10)/0.10$. The same conversion applies to costs, so value
and cost can be subtracted in one currency.

Finally, each scheme has its own best selectivity, so single-number comparisons
are made \textbf{best-versus-best}: each scheme at the $d$ that minimises its own
write amplification. Comparisons at matched $d$ are shown as curves, which are
the object of interest, and are never summarised as a range.

\subsection{What the collector can see, and what a label can add}
\label{sec:theory}
\label{sec:blackwell}

The theory starts from one observation. A collector that only chooses victims
decides using each block's valid count and age; it never looks at \emph{which}
pages are valid. Proposition~\ref{prop:perm} makes this precise, and everything
else follows from it.

\begin{proposition}[Permutation invariance]
\label{prop:perm}
Let a pool hold $b$ page slots and let $L_t$ be its count of valid pages. Any
collector whose information set contains \emph{only} per-pool valid counts and
pool ages --- greedy, $d$-choices, cost-benefit, garbage-ratio, FIFO reset ---
acts as a function on
$\mathcal{F}_C = \sigma(L_t,\ \text{ages})$, and this $\sigma$-algebra is
invariant under permutations of pages within a pool.
\end{proposition}

\begin{IEEEproof}
For pool $p$ with $b$ slots let $x_p\in\{0,1\}^{b}$ indicate which slots hold
valid pages, $L_p=\mathbf{1}^{\top}x_p$ its valid count, and $a_p$ its age. Let
$\pi=(\pi_p)$ be any family of within-pool permutations, $\pi_p\in S_b$, acting
by $x_p \mapsto x_p\circ\pi_p^{-1}$.

Such a permutation preserves the count,
$\mathbf{1}^{\top}(x_p\circ\pi_p^{-1})=\mathbf{1}^{\top}x_p=L_p$, and does not
touch $a_p$. Hence $(L,a)(\pi\cdot x)=(L,a)(x)$, so every
$\mathcal{F}_C$-measurable function is constant on the orbits of this action:
$\mathcal{F}_C\subseteq\mathcal{I}$, the $\sigma$-algebra of within-pool
permutation-invariant events.

Each rule named is a measurable function of $(L,a)$ alone: greedy is
$\arg\min_p L_p$; $d$-choices applies $\arg\min$ to a subset drawn independently
of $x$; cost-benefit maximises $(b-L_p)a_p/2L_p$; garbage-ratio maximises
$1-L_p/b$; FIFO orders by $a_p$. Each is therefore $\mathcal{I}$-measurable and
satisfies $A(\pi\cdot x)=A(x)$.

Finally, let $x,x'$ have identical per-pool counts and ages. Within each pool
the indicator vectors have equal weight, so some $\pi_p\in S_b$ carries one to
the other; with $\pi=(\pi_p)$ we get $x'=\pi\cdot x$ and $A(x')=A(x)$. The
collector cannot distinguish them.
\end{IEEEproof}

\begin{corollary}
\label{cor:residual}
Label value is exactly the information \emph{not} expressible as a
permutation-invariant statistic of per-pool occupancy.
\end{corollary}

In words: a cohort of pages placed together changes how $L_t$ evolves --- its
pool's valid count falls to zero at once --- and the collector sees that. The
same cohort scattered across pools contributes to each pool's count exactly as a
random set of pages would, and a collector that selects victims but does not
\emph{reorganise} data can never reassemble it. Structure that spans pools is
out of its reach by construction, not merely unexploited.

The hypothesis is that the information set is $(L,a)$, and two kinds of device
fall outside it. A firmware that tracks per-page history, such as the
migration-count classifier of Table~\ref{tab:roster}, uses a per-page signal and
could in principle learn membership from deallocation history, at a cost in
state and latency. The case for host-declared membership is therefore not that a
device can never infer it, but that the host holds it as a fact at no cost while
the device must pay to estimate it. Static wear
levelling~\cite{chang2007wearlevel,murugan2011rejuvenator} also falls outside,
because it relocates cold data on its own initiative: it reorganises rather than
selects, and could disturb or assemble cross-pool structure. Our propositions
cover the victim-selection component of a collector; what data migration does to
placement value is not examined here.

Corollary~\ref{cor:residual} says what is left for a label to carry. The second
step states the condition under which a label carries nothing at all.

\begin{proposition}[Conditional-information bound]
\label{prop:blackwell}
The residual information a label carries about invalidation time, given what the
collector already observes, is
\[
I(\Lambda;T\mid\mathcal{F}_C)
\]
for a label $\Lambda$ and invalidation time $T$. If the label is already
measurable with respect to what the collector observes --- that is, if
$\Lambda$ is $\mathcal{F}_C$-measurable --- then
$I(\Lambda;T\mid\mathcal{F}_C)=0$, and the label can add nothing the collector
does not already act on.
\end{proposition}

\begin{IEEEproof}
If $\Lambda$ is $\mathcal{F}_C$-measurable then $\Lambda$ is constant on each
atom of $\mathcal{F}_C$, so conditionally on $\mathcal{F}_C$ it is degenerate
and independent of $T$; hence $I(\Lambda;T\mid\mathcal{F}_C)=0$.
\end{IEEEproof}

Together the two steps give the paper's mechanism: a rate label is close to
$\mathcal{F}_C$-measurable, because occupancy already encodes rate, whereas a
membership label is not, because cross-pool membership is not a function of
$(L,a)$ at all.

The bound is deliberately weaker than a monotonicity statement. One might hope
that a richer collector always leaves less for a label to add, that is,
$I(\Lambda;T\mid\mathcal{F}_2)\le I(\Lambda;T\mid\mathcal{F}_1)$ whenever
$\mathcal{F}_1\subseteq\mathcal{F}_2$. Conditional mutual information is not
monotone in that way: with $\Lambda$ and $T$ independent fair bits and
$Z=\Lambda\oplus T$, the trivial $\mathcal{F}_1$ gives $I=0$ while
$\mathcal{F}_2=\sigma(Z)$ gives $I=1$ bit. A general ordering would need a
Blackwell-style sufficiency hypothesis~\cite{blackwell1953} that the collectors
studied here do not satisfy. The bound is also about information, not write
amplification: it does not say how much amplification a given amount of
information removes. The dependence of label value on the collector, and its
change of sign, are therefore established by measurement in
Section~\ref{sec:collector}, not predicted here. FIFO lies outside the
hypothesis altogether, because $\sigma(\text{age},\text{allocation order})$ and
$\sigma(L)$ are not nested, so its negative result is not a counterexample.

\subsection{How many classes, and how large a cohort}
\label{sec:regime}

Two further questions shape any labelling scheme: how many classes to use, and
whether the data being labelled are large enough for the collector to miss. One
proposition answers each.

\begin{proposition}[Cost-corrected value]
\label{prop:foc}
Write $\beta = \log_2 K$ for the number of label bits, and let
$V(\beta) = I(\beta \mid \mathcal{F}_C) - c(\beta)$, where $I$ is the
information a $2^{\beta}$-class labelling adds beyond $\mathcal{F}_C$ and $c$ is
the partitioning cost --- one write frontier per class, amortised spare
capacity, lost write locality, extra relocation. With $I$ increasing concave and
$c$ increasing convex \emph{in $\beta$}, $V$ has an interior maximiser
satisfying $I'(\beta^\ast) = c'(\beta^\ast)$.
\end{proposition}

\begin{IEEEproof}
$V'(\beta)=I'(\beta)-c'(\beta)$ with $I'>0$, $I''<0$, $c'>0$ and $c''>0$, so
$V'$ is strictly decreasing and has at most one zero. If $I'(0)>c'(0)$ and
$I'(\beta)<c'(\beta)$ for some $\beta$, that zero is an interior maximiser.
\end{IEEEproof}

We state this in $\beta$ rather than $K$ because that is where the measured cost
is clearly convex: on the measured grid the second divided differences of $c$
are $+0.84$ and $+1.54$ in $\beta$, against $+0.023$ and $+0.003$ in $K$, where
convexity is marginal (\S\ref{sec:kswept}); the grid $K \in \{2,4,8,16\}$ is
evenly spaced only in $\log_2$. Only the cost side is measured. Our instrument
returns $V=I-c$, so $I$ is recoverable only as $V+c$, and its concavity is an
assumption, not a finding; measuring $I$ alone would need a placement scheme with
no partitioning cost, which does not exist. The proposition explains an
observation otherwise easily dismissed: in class-count sweeps the profitable
$K^\ast$ falls as reclaim units grow (16, 8, 1 at unit sizes 10, 20, 40), which
is a first-order condition shifting, not an artifact. Fitting the cost
derivative and predicting $K^\ast$ at an unmeasured configuration would test the
condition rather than fit it; we have not done this.

\begin{proposition}[Scale ratio]
\label{prop:rho}
Let $\rho = C/P$ be a cohort's size $C$ relative to a reclaim unit's capacity
$P$. For $\rho < 1$ a cohort fits inside one pool, all of that pool's pages become invalid together,
and by Proposition~\ref{prop:perm} the collector obtains the structure for free
--- labels are redundant. For $\rho \gg 1$ the cohort must be assembled by the
host across pools or it is permanently unreachable. $\rho \approx 1$ is the
boundary.
\end{proposition}

\begin{IEEEproof}
If a cohort occupies at most one pool, its simultaneous invalidation drives that
pool's $L_p$ to zero, an event $\mathcal{F}_C$-measurable by
Proposition~\ref{prop:perm}; the collector reclaims it without relocation and
placement adds nothing. If a cohort spans $m>1$ pools and the reference layout
distributes its members across pools shared with other cohorts, then each pool's
$L_p$ trajectory is a sum over cohorts and no single cohort's invalidation brings
any pool's valid count to zero; by Proposition~\ref{prop:perm} the collector cannot identify the
membership that would, since that is not a function of $(L,a)$. Only a host that
co-locates the cohort changes the trajectory.
\end{IEEEproof}

Note that $\rho$ is a continuous size ratio, not the cohorts-per-unit count used
to admit traces below, which counts how many cohorts share a unit. The
proposition also classifies our own failed experiment. A RocksDB compaction trace
gave 1.00--1.68 cohorts per reclaim unit, median 1: its cohorts fit inside one
unit, $\rho < 1$, and the theory predicts what we found --- nothing for placement
to add --- though the trace also failed a steady-state check
(\S\ref{sec:workloads}).

\begin{conjecture}[Opposite scaling in unit size]
\label{conj:scaling}
As reclaim units grow, the two kinds of label scale in opposite directions. For
\emph{rate} labels, a larger pool contains more invalidation events, so the law
of large numbers flattens within-pool occupancy variance, the collector's lower
tail carries less, and label value \emph{rises}. For \emph{membership} labels, a
larger pool makes a cohort more likely to fit inside one pool ($\rho$ falls), so
the collector takes it free and label value \emph{falls}.
\end{conjecture}

\noindent The trend toward larger erase units would then make labels more and
less valuable at once, depending on the kind. Section~\ref{sec:grid} tests the
rate half.

\subsection{Workloads, admission rules and validation}
\label{sec:workloads}
\label{sec:anchors}

\textbf{Synthetic workloads.} The Rosenblum workload sends a fraction $r$ of
writes to a fraction $f$ of pages (the hot pages), with pages in each class
written equally often; it has rate structure but no cohort structure, and is the
control for every collector experiment. Membership experiments use cohort
workloads, in which groups of $C$ pages are created together and deleted
together (\S\ref{sec:t8}).

\begin{table}[t]
\caption{Admission rules for real traces, fixed before any trace was run
(\S S4). $\dagger$: membership experiments only.}
\label{tab:admission}
\centering
\footnotesize
\setlength{\tabcolsep}{4pt}
\begin{tabular}{ll}
\toprule
what it checks & rule \\
\midrule
steady state                      & overwrite ratio $\ge 3$ \\
something to eliminate            & $A_{\text{DWF}}\ge1.2$ at $d{=}24$ \\
exploitable structure             & oracle label beats DWF \\
deletions occur$^{\dagger}$       & within 20\% of creations \\
cohorts span units$^{\dagger}$    & $>1$ cohort per unit \\
\bottomrule
\end{tabular}
\end{table}

\textbf{Real traces.} A trace is admitted only if it passes the rules of
Table~\ref{tab:admission}. The criterion is not a formality: write order alone
moves cohorts-per-unit from 3.98 to 12.60 on synthetic data. Our RocksDB attempt
failed it --- 29 deletions across 40 SSTs, cohorts-per-unit 1.00--1.68 --- and
is reported, not used. Section~\ref{sec:trace} reports the one trace that
passed.

\begin{table}[t]
\caption{Validation of the simulator. Neither closed form is fitted to our data;
the last two rows reproduce the 2014 study~\cite{vanhoudt2014necessity} before
any new claim.}
\label{tab:validation}
\centering
\footnotesize
\setlength{\tabcolsep}{4pt}
\begin{tabular}{lll}
\toprule
check & expected & measured \\
\midrule
uniform-random, $u{=}0.90$   & $10.00$  & $9.984$ \\
FIFO fixed point             & $5.179$  & $5.241$ \\
three schemes, uniform       & equal    & $1.01\times$ \\
2014 ordering, $d{=}10$      & published & 5.84 / 2.99 / 2.19 \\
label-error crossover        & $\approx5\%$ & 6--10\% \\
\bottomrule
\end{tabular}
\end{table}

\textbf{Validation.} The collector axis is anchored at both ends by closed forms
(Table~\ref{tab:validation}; SWF on uniform writes, 12 passes, three seeds). The
FIFO fixed point is our own derivation, validated against measurement rather
than against an independent publication (\S S3). That the three schemes coincide
on uniform writes is the check that no scheme carries a label-independent bias,
as none can when there is nothing to separate.

%
%
\section{Label Value Depends on the Collector}
\label{sec:collector}

Section~\ref{sec:framework} predicts that a label is worth only what the
collector cannot already see, so its value should move with the collector. This
section measures how far. We vary victim selection alone on two synthetic
workloads (\S\ref{sec:sweep}), check that spare capacity does not stand in for
selectivity (\S\ref{sec:spare}), test the result across 48 configurations
(\S\ref{sec:grid}) and on a real trace (\S\ref{sec:trace}), and finally turn to
FIFO, under which the sign of label value reverses (\S\ref{sec:fifo}).

\subsection{Varying victim selection alone}
\label{sec:sweep}

\begin{table}[t]
\caption{Write amplification as only victim selection varies. Bold: each
scheme's best $d$.}
\label{tab:collector}
\centering
\begin{tabular}{lrrrc}
\toprule
$d$ & SWF & DWF & HCWF & winner \\
\midrule
\multicolumn{5}{l}{\emph{$r{=}0.9$, $f{=}0.05$}}\\
1      & 9.950 & 10.031 & 10.042 & tie \\
2      & 7.209 & 5.600 & 5.393 & HCWF \\
4      & 6.217 & 3.605 & 3.179 & HCWF \\
8      & 5.900 & \textbf{3.004} & 2.315 & HCWF \\
16     & 5.801 & 3.127 & \textbf{2.093} & HCWF \\
24     & \textbf{5.787} & 3.391 & 2.181 & HCWF \\
greedy & 5.820 & 4.721 & 3.866 & HCWF \\
\midrule
\multicolumn{5}{l}{\emph{$r{=}0.95$, $f{=}0.025$}}\\
1      & 9.943 & 10.014 & 10.007 & tie \\
2      & 7.291 & 5.374  & 5.327 & tie \\
4      & 6.350 & 3.115  & 2.995 & HCWF \\
8      & 6.016 & 2.207  & 1.941 & HCWF \\
16     & 5.916 & \textbf{2.030} & 1.563 & HCWF \\
24     & \textbf{5.899} & 2.139  & \textbf{1.502} & HCWF \\
greedy & 5.934 & 4.267  & 3.217 & HCWF \\
\bottomrule
\end{tabular}
\end{table}

\begin{figure}[t]
\centering
\includegraphics[width=0.7\columnwidth]{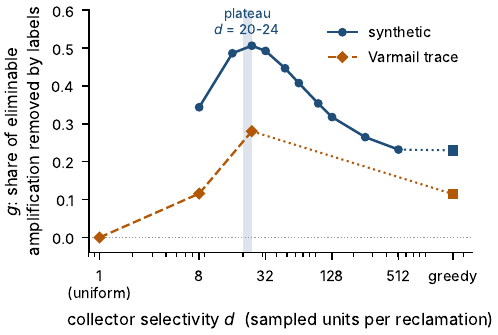}
\caption{What labels add over DWF, $g$, as victim selection varies. Circles:
synthetic workload; diamonds: Varmail trace.}
\label{fig:collector}
\end{figure}

Table~\ref{tab:collector} runs all three schemes on two Rosenblum workloads with
everything fixed except $d$: $b{=}32$, $S_f{=}0.10$, 12 passes, mean of five
seeds. A row is marked a \emph{tie} where DWF and HCWF differ by less than their
combined seed dispersion. Read best-versus-best, it gives the three accounts of
Table~\ref{tab:accounts}, none of them constant in the collector.
Figure~\ref{fig:collector} shows the same sweep as $g$. Greedy is not a value of
$d$, so its point is joined by a dotted line that implies nothing in between.

\textbf{Labels are worth most at intermediate selectivity.} $g$ rises from zero
under uniform selection to a peak and falls under greedy, and the peak is a
short, resolved plateau. With dispersion over five seeds, $g$ is $0.506$ at
$d{=}24$ and $0.505$ at $d{=}20$, against $0.486$ at $d{=}16$ and $0.492$ at
$d{=}32$ --- $3.4\sigma$ and $3.9\sigma$ below the peak, $\sigma$ being the
combined seed dispersion of the two points compared --- and more than $8\sigma$
below at $d{=}14$ and $d{=}48$. The ratio metric would not resolve this: at
$d{=}20$, $24$ and $32$ it gives values that lie within $0.7$ percentage points
of one another. We therefore report the plateau $d\in[20,24]$ rather than a
point.

\textbf{Each scheme has its own optimum, and the winner depends on the
collector.} On the first workload DWF is best at $d{=}8$, HCWF at $d{=}16$ and
SWF at $d{=}24$. Among collectors that read occupancy, HCWF is ahead of DWF
wherever the two can be told apart, and at $d{=}1$ (and $d{=}2$ on the second
workload) they tie. Under FIFO, which does not read occupancy, DWF wins by a wide
margin (\S\ref{sec:fifo}). The magnitudes move with the spare factor and the
workload; the reversal does not.

\textbf{Greedy is worse than an intermediate setting for every scheme.} Under
greedy all three are worse than at $d{=}16$ --- 5.820/4.721/3.866 against
5.801/3.127/2.093 --- though SWF only marginally, and HCWF at $d{=}16$ is
$1.85\times$ better than HCWF under greedy. This extends the published
observation that greedy is not optimal for frontier
schemes~\cite{vanhoudt2013hotcold}: an aggressive collector reclaims blocks that
would have lost more of their valid pages had it waited, which hurts most the
schemes whose purpose is to let whole blocks become invalid.

\textbf{The $d{=}1$ row is an identity, not a measurement.} Under uniform
selection the collector cannot exploit any arrangement, so the label returns
nothing, and HCWF and DWF both hold two frontiers, so it costs nothing. The three
schemes nearly coincide, as they must, and we draw nothing from that row.

For FDP, this means identical labels on identical workloads should return
differently on different drives, which is the instability that cross-device
characterisation reports~\cite{warp2026}. We offer collector selectivity as
\emph{one candidate mechanism alongside} the causes it names, not as an
explanation of its findings.

\subsection{Spare capacity does not substitute for selectivity}
\label{sec:spare}

\begin{table}[t]
\caption{Label value in the ratio metric and in $g$, against selectivity and
spare factor ($r{=}0.9$, $f{=}0.05$).}
\label{tab:g}
\centering
\begin{tabular}{lrr|lrr}
\toprule
\multicolumn{3}{c|}{selectivity ($S_f{=}0.10$)} & \multicolumn{3}{c}{spare ($d{=}24$)} \\
$d$ & ratio & $g$ & $S_f$ & ratio & $g$ \\
\midrule
8      & 22.9\% & 0.344 & 0.05 & 40.4\% & 0.489 \\
16     & 33.1\% & 0.486 & 0.10 & 35.7\% & 0.506 \\
24     & 35.7\% & \textbf{0.506} & 0.20 & 28.1\% & 0.560 \\
32     & 35.5\% & 0.492 & 0.30 & 21.1\% & \textbf{0.625} \\
64     & 30.7\% & 0.407 & & & \\
greedy & 18.1\% & 0.230 & & & \\
\bottomrule
\end{tabular}
\end{table}

An obvious objection is that spare capacity and selectivity do the same job:
both make blocks holding little valid data easier to find. If so, a spare sweep should
reproduce the curve of Fig.~\ref{fig:collector}. Table~\ref{tab:g} tests this
on the first workload ($b{=}32$, 12 passes, five seeds), in both metrics.

\textbf{Along selectivity, the two metrics agree in shape.} In $g$ the optimum is
interior and in the same place, and the fall to greedy is steeper than the ratio
suggests: $0.506$ to $0.230$, a factor of $2.2$, against $35.7\%$ to $18.1\%$.
The second workload behaves the same way: at $r{=}0.95$ the ratio runs $3.8\%$,
$12.1\%$, $23.0\%$, $29.8\%$ and $40.4\%$ at $d{=}4$ to $64$ and falls to
$24.6\%$ under greedy, while $g$ runs $0.057$ to $0.617$ and falls to $0.321$.

\textbf{Along spare, they disagree.} In the ratio, label value falls as
over-provisioning grows, which reads as spare substituting for labels. In $g$ it
\emph{rises}: $0.489$, $0.506$, $0.560$, $0.625$ at $S_f = 0.05$ to $0.30$ at
$d{=}24$, with the same direction at $d{=}8$ ($0.335 \to 0.450$) and under
greedy ($0.249 \to 0.436$). The fall in the ratio is compression: both
amplifications approach one, so any relative measure must shrink. The corrected
reading is that spare removes the collector's own margin --- with ample spare a
block holding little valid data is always at hand, so selection has little left to do --- and
what remains is increasingly the label's to take. \textbf{Buying flash does not
substitute for labelling; it makes labelling a larger share of a smaller
residual.}

So spare and selectivity are not one axis. There is a price between them rather
than an identity: were they one axis, the capacity equivalent of
\S\ref{sec:capeq} would be a constant, and it is not. This also says that the
collector's residual has at least two components, one from pressure and one from
search, which makes the theory of Section~\ref{sec:framework} incomplete rather
than wrong.

\subsection{Robustness across 48 configurations}
\label{sec:grid}

The results so far rest on two workloads at one device size and one spare
factor. We swept four workloads, four reclaim-unit sizes ($b$ from 16 to 128
pages) and three spare factors --- 48 cells --- over
$d\in\{4,8,16,24,32,48,64\}$ and greedy, five seeds each, at 12 passes. The
verdict rule was fixed before the run (\S S4): the shape holds in a cell if the
best $d$ is interior to the grid ($4<d^\ast<64$), and both the rise from $d{=}4$
and the fall to greedy exceed $2\sigma$ (combined seed dispersion of the two
points compared). At this run length every cell has $A_{\text{base}}\ge1.5$, so
all 48 are reportable.

\textbf{It holds in 47 of the 48 cells}. Over those 47, the rise from $d{=}4$
runs from $15.0\sigma$ to $74.4\sigma$ (median $33.0\sigma$) and the fall to
greedy from $12.5\sigma$ to $141.8\sigma$ (median $48.2\sigma$), with
$d^\ast\in\{8,16,24,48\}$. The remaining cell ($r{=}0.95$, $f{=}0.025$,
$b{=}32$, $S_f{=}0.10$) fails, and we report it as a failure. Its best $d$ is 64,
the top of the grid: $g$ is $0.6156$ at $d{=}48$ and $0.6174$ at $d{=}64$, a tie
at $0.4\sigma$, and falls by $34\sigma$ to greedy. Rise and fall are both
present; what the test cannot establish is that the peak lies inside the grid.
Every other cell of that workload peaks at $d{=}48$, one step below the edge, so
the grid we fixed in advance was barely wide enough for it.

\textbf{The optimum belongs to the workload, not the geometry.} Across an
eightfold range of reclaim-unit size, $d^\ast$ moves by at most one grid step
within a workload, while across workloads it moves from 8--16 ($r{=}0.8$,
$f{=}0.10$) to 48--64 ($r{=}0.95$, $f{=}0.025$). The optimum does not transfer
between workloads, but it is stable against the device parameter we varied most
widely. This also means that one pre-registered expectation, that the optimum
would shift with unit size, was not met.

\textbf{The spare result replicates.} The best $g$ rises with $S_f$ in all four
workloads at every unit size, with one tie ($r{=}0.95$, $b{=}128$: $0.637$ at
both $S_f{=}0.07$ and $0.10$) --- for instance $0.318 \to 0.330 \to 0.386$ at
$r{=}0.8$, and $0.493 \to 0.506 \to 0.560$ at $r{=}0.9$, $f{=}0.05$, both at
$b{=}32$.

\textbf{The rate half of Conjecture~\ref{conj:scaling} holds at moderate spare.}
Holding workload and spare fixed and growing units from 16 to 128 pages, the best
$g$ rises by more than $2\sigma$ (combined seed dispersion) in 7 of the 12
workload--spare pairs and falls in none; the 5 flat pairs are all four at
$S_f{=}0.20$ and $r{=}0.8$ at $S_f{=}0.10$. That fits the conjectured mechanism:
a larger pool flattens occupancy variance, and so does more spare, so when spare
is plentiful the larger pool has little left to add. The membership half is not
tested here.

\subsection{A real trace}
\label{sec:trace}

\begin{table}[t]
\caption{Write amplification and $g$ on the Varmail trace as victim selection
varies, with oracle labels.}
\label{tab:trace}
\centering
\footnotesize
\setlength{\tabcolsep}{4.5pt}
\begin{tabular}{lrrrrc}
\toprule
collector & SWF & DWF & HCWF & $g$ & winner \\
\midrule
uniform  & 10.007 & 10.003 & 10.004 & 0.000  & tie \\
$d{=}8$  & 2.539  & 2.168  & 2.033  & 0.116  & HCWF \\
$d{=}24$ & 2.153  & \textbf{1.704} & \textbf{1.507} & \textbf{0.280} & HCWF \\
greedy   & 3.061  & 2.639  & 2.452  & 0.114  & HCWF \\
FIFO     & 7.333  & 7.333  & 7.634  & $-0.048$ & DWF \\
\bottomrule
\end{tabular}
\end{table}

Every result above uses a synthetic workload. To test whether the shape
survives a real one, we ran the same sweep on the Varmail block trace released
with the system discussed in \S\ref{sec:dogi}~\cite{dogi2026,dogitraces}: the
Filebench mail-server workload, recorded beneath a log-structured file system.
The trace records only 4\,KiB writes, with no discards, so a page the file
system has freed stays valid until overwritten, as on a drive without TRIM. The
device is sized to the pages the trace writes, with $b{=}32$ and $S_f{=}0.10$.

A trace carries no labels, so we used the strongest rate label it admits: an
oracle that marks as hot the pages with the most writes over the whole trace.
The hot fraction was chosen once, at $d{=}24$, from
$\{0.002,0.005,0.01,0.02,0.05,0.10,\dots,0.30\}$ --- the best, 0.01, is interior
to that grid --- and then held fixed while the collector varied, as in every
synthetic experiment. Varmail's working set grows throughout the trace: prefixes
of 2M, 10M and 50M writes touch 1.00M, 3.83M and 16.3M distinct pages. The first
two fail the steady-state test (overwrite ratios 1.99 and 2.61 against 3). The
50M-write prefix passes all three tests: overwrite ratio 3.07,
$A_{\text{DWF}}{=}1.704$, and an oracle label (hot fraction 0.01) that lowers
amplification to 1.507.

\textbf{Both claims hold, at about half the magnitude}
(Table~\ref{tab:trace}, Fig.~\ref{fig:collector}). What labels add rises from
nothing under uniform selection to $g{=}0.280$ at $d{=}24$ and falls to $0.114$
under greedy: an interior optimum, as on every synthetic workload. Under FIFO the
same labels cost more than they return, $g{=}-0.048$, and the label-free scheme
wins. The peak is a little over half the synthetic headline of $0.506$; since
$d^\ast$ and the size of $g$ are workload properties (\S\ref{sec:grid}), we read
this as agreement in shape, not as a measurement of the effect's size on real
systems.

The result is stable within the trace. Where the collector draws random numbers
we averaged three seeds (seed s.d.\ $\le0.007$ in write amplification and
$\le0.001$ in $g$); greedy, FIFO and SWF draw none. The three seeds agree to
$0.001$ in $g$, and the next 50M writes --- a second window of the same
workload, not an independent trace --- reproduce every $g$ within $0.002$ and
every write amplification within $0.2\%$. Two limits apply. It is one workload,
and the reversal under FIFO is a $4\%$ difference, far smaller than the
synthetic one (\S\ref{sec:fifo}), though it recurs in the second window. The
trace tests the collector claim with rate labels; it says nothing about
membership labels (\S\ref{sec:t8}).

\subsection{A blind collector reverses the sign}
\label{sec:fifo}

FIFO reclaims in allocation order without looking at occupancy. It is not the
$d{=}1$ limit of $d$-choices: uniform-random selection at least samples the
occupancy distribution, while FIFO's order follows age and is therefore
\emph{anti}-correlated with occupancy. Its information set,
$\sigma(\text{age},\text{allocation order})$, neither contains nor is contained
in $\sigma(\text{per-pool counts})$, so Proposition~\ref{prop:blackwell} makes
no prediction about it.

\begin{table}[t]
\caption{HCWF under FIFO against run length ($r{=}0.9$, $f{=}0.05$).}
\label{tab:fifo}
\centering
\begin{tabular}{rrrr}
\toprule
passes & mean & s.d. & $\Delta$mean \\
\midrule
8   & 15.713 & 2.607 & --- \\
16  & 16.552 & 2.898 & $+5.3\%$ \\
32  & 17.259 & 1.518 & $+4.3\%$ \\
64  & 15.970 & 1.104 & $-7.5\%$ \\
128 & 16.469 & 0.308 & $+3.1\%$ \\
\bottomrule
\end{tabular}
\end{table}

On uniform writes FIFO is well behaved and matches its closed form
(\S\ref{sec:workloads}). On the skewed workload HCWF oscillates, and we ran it
long enough to say what kind of oscillation it is (Table~\ref{tab:fifo}: the
same five workload seeds at each run length, so its 32-pass row is the FIFO
column of Table~\ref{tab:accounts}). The mean rises and falls with no trend,
while the seed dispersion collapses by a factor of about eight. A divergent
process would show a trend and a dispersion that grows or holds; this shows
neither. \textbf{It is a bounded limit cycle}, and we
report it as a value with an interval: over 32 passes and five workload seeds,
$A_{\text{HCWF}} = 17.3 \pm 1.5$ against a stable $A_{\text{DWF}} = 6.98$,
giving $g = -1.72$. By 128 passes HCWF has settled to $16.5\pm0.3$. Shorter
runs, at which the sequence has not yet revealed its bound, would suggest no
converged value at all.

\textbf{The reversal is robust.} FIFO draws no random numbers, so we varied
what it is applied to: on both synthetic workloads, at $N{=}300$, 600 and 1200
blocks, over five workload seeds of 32 passes each, DWF beats HCWF in every run,
by a factor of $1.3$ to $2.8$ per run ($1.5$ to $2.6$ averaged over seeds), and
doubling the run length leaves the sign unchanged. Across these six
configurations labels under a blind collector do not merely fail to help: they
cost between $0.55$ and $1.84$ times the eliminable amplification. Host
separation buys nothing under FIFO either: SWF settles at $6.93$--$6.94$ at
$N{=}300$, $600$ and $1200$ (seed s.d.\ at most $0.003$), marginally below DWF's
$6.96$--$7.03$. A blind collector gains nothing from separating relocations and
loses from separating by label.

\textbf{A blind collector also pays for a split that carries no information.}
Splitting a homogeneous workload into two classes that differ only in their
label, so that the label carries nothing, the same split costs $-0.0\%$ under
$d{=}24$ and $+11.1\%$ under FIFO, with labels, balance and device fixed (\S S1).
A selective collector routes around a poor split; a blind one pays for it.

\textbf{The instability belongs to the labelled scheme.} Both label-free
schemes are stable under the same collector and workload. HCWF differs in
splitting the stream by label and in relabelling blocks on reclaim, so relocated
data re-enter the frontiers that serve host writes; we have not isolated which
of the two produces the cycle, nor recorded whether the free-block reserve is
exhausted during it. With the queue semantics fixed rather than swept
(\S\ref{sec:method}), our magnitudes are not the worst case.

%
%
\section{What Labels Are Worth, and Which Labels Pay}
\label{sec:labels}

Section~\ref{sec:collector} shows that the collector sets how much a label can
be worth. A host or operator deciding whether to label needs more than that.
This section prices a label in spare capacity (\S\ref{sec:capeq}), asks what
remains at realistic label error (\S\ref{sec:errcap}), tests whether more classes
help (\S\ref{sec:kswept}), and compares rate labels with membership labels, the
one kind the theory says the collector cannot replace
(\S\ref{sec:t8}, \S\ref{sec:kappa}).

\subsection{The capacity equivalent of a label}
\label{sec:capeq}

\begin{table}[t]
\caption{Capacity equivalent of a two-class label at $S_f{=}0.10$.}
\label{tab:capeq}
\centering
\small
\setlength{\tabcolsep}{4.5pt}
\begin{tabular}{lrrrr}
\toprule
comparison & $d$ (HCWF) & $d$ (DWF) & $\Delta S/S$ & s.d. \\
\midrule
matched        & 1  & 1  & $+0.6\%$  & 0.7 \\
matched        & 8  & 8  & $+44.6\%$ & 0.9 \\
matched        & 24 & 24 & $+83.2\%$ & 0.8 \\
best-vs-best   & 16 & 10 & $\mathbf{+64.7\%}$ & 0.4 \\
\bottomrule
\end{tabular}
\end{table}

Table~\ref{tab:capeq} gives the capacity equivalent $\Delta S/S$
(\S\ref{sec:metric}) on the first workload. \emph{Matched} rows run both schemes
at the same $d$; the \emph{best-versus-best} row runs each at the $d$ that
minimises its own write amplification. The spare grid is resolved to one point
($S_f$ from 0.05 to 0.25 in steps of 0.01), with three seeds and 32 passes; every
entry moves by less than two points from 24 passes, and seed dispersion is under
one point.

\textbf{Which comparison is fair.} Matched selectivity is not: at $d{=}24$ DWF is
far from its own optimum, so the matched $+83.2\%$ flatters the labels. Searching
both schemes over the same one-step grid ($d\in\{4,6,8,10,12,16,20,24,32\}$)
puts HCWF at $d{=}16$ and DWF at $d{=}10$, giving $+64.7\%\pm0.4$. We take that
as the paper's claim. HCWF's exact $d$ matters little near its optimum
($+63.0\%$ at $d{=}12$, $+60.6\%$ at $d{=}20$). DWF's optimum is $d{=}10$ on this
finer grid against $d{=}8$ on the coarser grid of Table~\ref{tab:collector};
HCWF's is $d{=}16$ on both.

So at a 10\% spare factor on this workload, a label is worth roughly 65\%
additional over-provisioning: matching the labelled scheme without labels would
take a drive with about 16.5\% spare rather than 10\%. Under uniform-random
selection the same label is worth $+0.6\%\pm0.7$ --- nothing. The figure is not
a constant: it depends on selectivity, spare factor and workload, and quoting it
without all three would repeat the reporting practice we criticise. These
figures need a long run: HCWF's amplification is still rising after a few
passes (\S S5).

\subsection{The capacity equivalent under realistic label error}
\label{sec:errcap}

\begin{figure}[t]
\centering
\includegraphics[width=0.7\columnwidth]{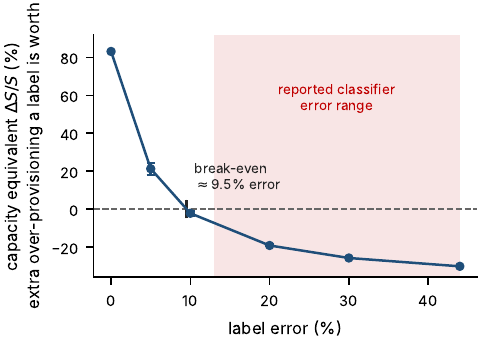}
\caption{Capacity equivalent of a two-class label against label error. Shaded:
error range reported for learned lifetime classifiers.}
\label{fig:errcap}
\end{figure}

Every figure above assumes perfect labels. Deployed lifetime classifiers are
not perfect: the most recent system reports 56--84\% accuracy for its own two
class types, and 76--87\% for the prior schemes it evaluates~\cite{dogi2026}. We
take 13--44\% as the span of \emph{reported} error; we know of no field
measurement of classifier accuracy in production. Figure~\ref{fig:errcap}
repeats the matched $d{=}24$ measurement with a fraction of labels flipped at
random (values in \S S6).

\textbf{The capacity equivalent crosses zero between 5\% and 10\% error, before
the reported range begins.} Linear interpolation between the bracketing points
places the crossing at $\approx 9.5\%$ error, that is $\approx 90\%$ accuracy;
at 10\% error the value is already $-2.2\%\pm1.2$. Every error rate reported for
these classifiers lies on the negative side. At 20\% error the labelled scheme
is worth $-19.2\%$ over-provisioning: an operator would prefer the label-free
collector \emph{and} 19\% less flash. At the pessimistic end of the reported
range it is $-30.2\%$. Because matched $d{=}24$ flatters the labels
(\S\ref{sec:capeq}), these negative values are conservative.

The $+83.2\%$ at zero error is therefore not what labelling is worth; it is the
ceiling of what it could be worth if classification were solved. The collector
decides how much is available to win; the classifier decides whether any of it
is collected.

\subsection{More classes buy almost nothing}
\label{sec:kswept}

If two classes are not enough, perhaps more are. Proposition~\ref{prop:foc} says
the answer turns on a cost that grows with the number of classes, so we first
measure that cost on its own, then the net value.

\begin{table}[t]
\caption{Partitioning cost $c(K)$ of $K$ information-free label classes.}
\label{tab:cost}
\centering
\begin{tabular}{llrrrr}
\toprule
$S_f$ & $d$ & $K{=}2$ & $K{=}4$ & $K{=}8$ & $K{=}16$ \\
\midrule
0.10 & 24     & 0.6\% & 1.8\% & 4.5\% & 10.4\% \\
0.10 & greedy & 0.7\% & 2.3\% & 5.0\% & 10.9\% \\
0.20 & 24     & 0.3\% & 0.9\% & 1.9\% & 4.5\% \\
0.20 & greedy & 0.3\% & 1.0\% & 2.0\% & 4.5\% \\
\bottomrule
\end{tabular}
\end{table}

\textbf{The cost of a partition.} We measure the cost by giving the labels no
information at all: a homogeneous workload split into classes that differ only
in their label. With two classes the cost is indistinguishable from zero,
$-0.2\%$ to $+0.4\%$ across four spare factors and three selectivities, because
HCWF and DWF both hold two frontiers and HCWF pays for no extra one. With $K$
data frontiers plus one for relocated data, the cost grows
(Table~\ref{tab:cost}; uniform writes, 12 passes, three seeds). It is increasing
and convex in the number of label bits, by a factor of $2.77$, $2.59$ and $2.29$
per doubling of $K$, which is all Proposition~\ref{prop:foc} needs (second
divided differences $0.84$ and $1.54$); in $K$ itself convexity is marginal
($+0.023$ and $+0.003$). It falls by $2.2\times$ to $2.3\times$ when spare
doubles from $0.10$ to $0.20$, and depends little on selectivity: greedy and
$d{=}24$ differ by at most half a point.

The cost is not write imbalance. Sweeping an information-free split from 50\% to
2\% of the address space costs at most $+0.5\%$ under $d{=}24$, and the thinnest
class still spans about eleven reclaim units, so this is not an artifact of a
class too small to separate. It is not a per-frontier capacity quota either: at
this device size no frontier reaches its equal share, the most any holds being
76 blocks of a 200-block quota at $K{=}2$, and capped and uncapped runs agree
exactly (\S S1). What $c$ measures is therefore the time each frontier takes to
fill and the write locality lost by splitting. The simulator shares one free
pool among all frontiers, which is not a capacity rule but the \emph{absence} of
one; no device can implement it and it is the choice most favourable to labelled
placement, so \textbf{these costs are a lower bound}.

\begin{table}[t]
\caption{Net value of $K$ nested label classes, after partitioning cost.}
\label{tab:net}
\centering
\begin{tabular}{rrrr}
\toprule
$K$ & vs DWF & vs SWF & frontiers \\
\midrule
1  & $0.0\%$ & $5.0\%$ & --- \\
2  & $+0.5\%$ & $+5.5\%$ & 3 \\
4  & $\mathbf{+1.3\%}$ & $\mathbf{+6.2\%}$ & 5 \\
8  & $+0.2\%$ & $+5.2\%$ & 9 \\
16 & $-7.2\%$ & $-1.8\%$ & 17 \\
\bottomrule
\end{tabular}
\end{table}

\textbf{The net value.} Table~\ref{tab:net} uses a workload whose traffic halves
across 16 equal address classes, so coarser labels are unions of finer ones and
the partitions are nested: the $K{=}16$ labels determine the $K{=}8$ labels, and
a fall in value cannot be blamed on a change of boundaries ($S_f{=}0.10$,
$d{=}24$, 6 passes, three seeds). Each entry is the reduction in write
amplification of the $K$-class partition against the named baseline, minus the
cost $c(K)$ measured separately; the entries are already net. $K^\ast{=}4$
against both baselines, but the magnitude differs sharply: $+1.3\%$ against DWF
but $+6.2\%$ against SWF. That is the paper's thesis in one line --- the same
labelling looks five times more valuable against the weaker reference. $K{=}1$
reproduces DWF exactly, which anchors the family.

\textbf{The profitable class count belongs to the labelling.} Three
constructions on related workloads give $K^\ast{=}2$, $K^\ast{=}4$, and no class
count worth more than about 1\%, differing only in how the classes are cut
(\S S2). A paper that reports $K^\ast$ without its partition rule has therefore
not reported enough to be reproduced. All three share one feature: the benefit
is largest at small $K$ and gone by $K{=}16$.

\textbf{With a correct partition, rate labels add about 1\%.} One of those
sweeps cut the address space into $K$ equal pieces. On this workload the hot
pages are 5\% of the address space, so a two-way address split produced a class
that was only 10\% hot --- a poor lifetime partition. Relabelling by \emph{rate
quantile} gives a class that is 100\% hot, and the benefit nearly vanishes:
$+0.3\%$, $+0.3\%$ and $+1.0\%$ at $K{=}2,4,8$ against DWF, with seed s.d.\
$0.3$--$0.5$ points (12 passes, three seeds). The earlier interior optimum was an
artifact: with a partition that mixed lifetimes, extra frontiers partly undid
the damage, which read as benefit. Once DWF has separated host writes from
relocations, further rate-based separation adds at most about 1\% at any class
count we measured, \emph{even with a perfect partition}. This is the theory's
prediction on the class-count axis: a searching collector already infers what
rate labels encode. ($\mathrm{KWF}$ extends DWF rather than HCWF, so these
figures are not comparable with the two-class ones above; \S S2.)

\subsection{Membership labels at deployable handle counts}
\label{sec:t7power}
\label{sec:t8}

\begin{figure}[t]
\centering
\includegraphics[width=0.7\columnwidth]{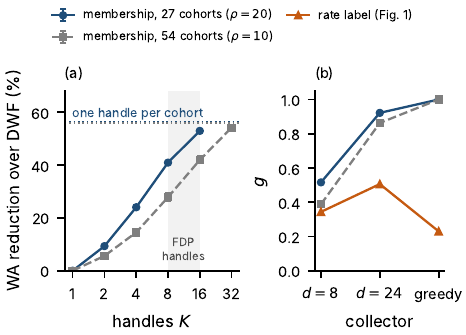}
\caption{Membership labels. (a) Write-amplification reduction over DWF against
the number of handles ($d{=}24$); dotted: one handle per cohort; shaded: handle
counts of commercial FDP devices. (b) $g$ against the collector, with one handle
per cohort, beside the rate label of Fig.~\ref{fig:collector}.}
\label{fig:handles}
\end{figure}

Rate labels predict; membership labels assert. A membership label tells the
device which pages belong to one object and will be deleted together --- a fact
the host holds and, by Proposition~\ref{prop:perm}, the collector cannot
reconstruct once the pages are scattered. Proposition~\ref{prop:rho} says such
labels matter when cohorts are larger than a reclaim unit. We test them on
cohort workloads with $\rho = C/P = 20$ and $10$ (cohorts of 640 and 320 pages,
32-page units), over five workload seeds.

The ideal assigns one handle per cohort --- 27 and 54 cohorts in our two
configurations, of which 16 and 32 are live at once --- far beyond what any
device offers. With $K$ handles we assign cohort $c$ to handle $c \bmod K$.
That rule uses only which cohort a page belongs to, not when the cohort will
die, so it is a conservative stand-in for a host that groups cohorts by expected
death. Fig.~\ref{fig:handles}(a) gives the result at $d{=}24$: $+24.0\%$,
$+40.9\%$ and $+52.9\%$ at $K{=}4$, $8$ and $16$ handles with 27 cohorts,
against $+56.4\%$ with one handle per cohort, and $+14.5\%$, $+27.8\%$ and
$+41.9\%$ with 54, against $+55.8\%$ (seed s.d.\ at most $1.4$ points). The
gain rises steadily with $K$, and with 54 cohorts 32 handles reach $+54.1\%$;
16 is not a special point.

\textbf{At sixteen handles --- within what commercial FDP devices expose ---
membership labels capture $94\%$ and $75\%$ of what one handle per cohort
achieves.} This is the paper's one positive result about labels, and it survives
the handle limit. It contrasts sharply with rate labels, which add at most about
1\% beyond DWF at any class count, while membership labels at a realisable $K$
remove more than half the eliminable amplification. The difference is not how
many classes, but what the label carries: a fact rather than a prediction.

\textbf{Unlike rate labels, membership labels gain from a better collector.}
Fig.~\ref{fig:handles}(b) repeats the one-handle-per-cohort case under other
collectors. $g$ is $0.516$ at $d{=}8$, $0.922$ at $d{=}24$ and $1.000$ under
greedy with 27 cohorts ($0.388$, $0.863$ and $1.000$ with 54). Under greedy no
reclaimed block holds valid data: each cohort fills its own blocks and dies
whole, and greedy always finds such a block. The rate label of
Fig.~\ref{fig:collector} instead peaks at intermediate selectivity and falls
under greedy. This is the contrast Propositions~\ref{prop:perm}
and~\ref{prop:blackwell} predict: a searching collector can infer rate from
occupancy, so it competes with a rate label, but it cannot infer membership, so
a better search only lets a membership label pay more. FIFO is again the
exception: blind to occupancy, it cannot use membership either, and one handle
per cohort costs $-12.6\%$ and $-41.0\%$ (\S S6).

Two limits apply. The handle assignment ignores death order; a host that knows
death order could do better, but an estimate of it would carry error, which is
what destroys label value (\S\ref{sec:errcap}). And these are synthetic
cohorts, not a trace.

\subsection{A workload statistic that cannot work}
\label{sec:kappa}

If label value could be predicted from a workload statistic, an operator could
screen workloads before labelling them. Proposition~\ref{prop:perm} says a
statistic of \emph{within}-pool invalidation structure cannot do this, because
that structure is visible to the collector already. We built one and confirmed
it. $\kd$ is the largest fraction of a reclaim unit's blocks invalidated within
a window of $k$ fill durations, under a no-placement reference. Once the
available gain is known it predicts little: its partial $R^2$ on the label-error
crossover given zero-error gain is $0.151$ over 15 workloads, and in a matched
design (180 candidates screened, 83 workloads in 26 groups of equal gain) it
explains $R^2 = 0.167$. Its window barely matters --- $\kd(k{=}1)$ and
$\kd(k{=}50)$ correlate at $0.86$ to $0.95$ across three sweeps --- and it tracks
the invalidated fraction, $\operatorname{corr}(\kd, 1-u) = 0.97$ on the
concurrency sweep: a permutation-invariant count, exactly what
Proposition~\ref{prop:perm} says the collector already has (\S S7). A useful
statistic would instead measure how cohorts interleave \emph{across} pools,
which the collector cannot reach; cohorts-per-unit is a first attempt.

%
%
\section{Conclusion and Discussion}
\label{sec:deliver}

Placement benefit divides into three terms: the coarse separation a device
obtains without labels, its victim-selection policy, and the labels themselves.
The middle term is of the same order as the other two, is rarely reported, and
under FDP the host neither chooses nor observes it. Labels are worth most under
a collector of intermediate selectivity, less under greedy, nothing under
uniform selection, and less than nothing under FIFO, so which scheme to deploy
is itself a function of the collector. Prior results are not wrong, but they do
not transfer unless the collector is reported.

\subsection{Implications}
\begin{itemize}
\item \textbf{Report the collector, and the free baseline.} State the
victim-selection policy and its setting, and measure against DWF as well as
SWF. The same labels look five times more valuable against the weaker reference
(Table~\ref{tab:net}), and the winner changes with the collector
(Table~\ref{tab:collector}). Under FDP the host cannot even select that
baseline: separating host writes from relocations is a firmware action, which
ZNS exposes and FDP does not.

\item \textbf{Assert membership; do not predict lifetime.} Membership and
deletion events lie outside what the collector observes, and the host holds them
as facts rather than predictions. This suggests one more hint beside the FDP
reclaim unit handle --- \emph{these writes belong to object $X$, which dies at
event $Y$} --- which is new information, not a new algorithm. Use it where
cohorts span many reclaim units ($\rho \gg 1$): SST groups, container writable
layers, KV-cache prefixes. Below $\rho < 1$ the collector wins for free.

\item \textbf{Estimate the hidden variable.} Under FDP the host sees only
telemetry. An estimator of effective selectivity from observables --- the
distribution of valid fractions at reclaim, relocation latency, the derivative
of write amplification with respect to host writes --- would let an operator
predict what a labelling scheme is worth on a drive before deploying it.

\item \textbf{Decide in capacity, not in ratio.} Use $\Delta S/S$, and compare
it with the price of the flash it displaces. A value above zero by more than its
measurement error is necessary but not sufficient: the economic decision also
needs the price per gigabyte and the drive's warranted lifetime. The shape is
measured here; the coefficients for a real drive need its collector.
\end{itemize}

\subsection{Limitations}
We use simulation because it is the only way to hold the workload, the labels,
the device and the spare capacity fixed while varying victim selection alone;
hardware would test magnitudes and vendor policies rather than the mechanism,
which is what the collector can see, and every victim-selecting collector in the
class we model sees only valid counts and ages. 


\medskip
\noindent What a label is worth is not a property of the label. Until an
evaluation says which collector it ran under, it has not said what it measured.

\bibliographystyle{IEEEtran}
\bibliography{synchrony}

\section*{Biographies}

\noindent\textbf{Nannan Zhao}
is an Associate Professor at Northwestern Polytechnical University. Her research interests include distributed storage systems, flash memory, and memory architectures.
\par\vspace{0.3\baselineskip}

\noindent\textbf{Jiameng Zhang}
received the B.S. degree in computer science from Northwestern Polytechnical University in 2025, where she is pursuing the M.S. degree. Her research interests include cloud storage and distributed systems.
\par\vspace{0.3\baselineskip}

\noindent\textbf{Taoyu Zhong}
is pursuing the B.S. degree in computer science at Northwestern Polytechnical University. His research interests include cloud storage and distributed systems.
\par\vspace{0.3\baselineskip}

\noindent\textbf{Bo Wang}
received the B.Eng. degree from Northwestern Polytechnical University in 2026, where he is pursuing the M.S. degree in computer science. His research interests include computer systems and storage systems.
\par\vspace{0.3\baselineskip}

\noindent\textbf{Hao Wang}
received the B.S. degree from Northwestern Polytechnical University in 2023, where he is pursuing the M.S. degree in computer science and technology. His research interests include cloud computing and resource management.
\par\vspace{0.3\baselineskip}

\noindent\textbf{Zhijie Huang}
received the Ph.D. degree from Huazhong University of Science and Technology in 2016. He is an Associate Professor at Northwestern Polytechnical University. His research interests include coding theory, storage systems, and blockchain.
\par\vspace{0.3\baselineskip}

\noindent\textbf{Shujie Han}
received the B.Eng. degree from Northwestern Polytechnical University in 2017 and the Ph.D. degree from The Chinese University of Hong Kong in 2021. She is an Associate Professor at Northwestern Polytechnical University. Her research interests include storage reliability and AI for storage.
\par\vspace{0.3\baselineskip}

\noindent\textbf{Xiaonan Zhao}
received the Ph.D. degree in computer science and technology in 2010. She is an Associate Professor at Northwestern Polytechnical University. Her research interests include distributed file systems and cloud storage optimization.
\par\vspace{0.3\baselineskip}

\noindent\textbf{Xiao Zhang}
is a Professor at Northwestern Polytechnical University. His research interests include storage systems, computer networks, and AI systems.
\par

\clearpage


\setcounter{section}{0}
\setcounter{subsection}{0}
\setcounter{table}{0}
\setcounter{figure}{0}
\setcounter{equation}{0}

\renewcommand{\thesection}{S\arabic{section}}
\renewcommand{\thetable}{S\arabic{table}}
\renewcommand{\thefigure}{S\arabic{figure}}
\renewcommand{\theequation}{S\arabic{equation}}

\providecommand{\kd}{\kappa}

\section*{Supplemental Material}

\noindent
This material supports the paper of the same name and is referenced
from it as \S S1--\S S7. It contains supporting measurement, the derivation of
one fixed point, the pre-registered verdict rules, and the record of corrections
that measurement forced on earlier drafts. Nothing here is required to follow
the paper's argument; everything here is required to check it. Section numbers
such as Sec.~IV-C refer to the paper.

\section{What the partitioning cost is not}
\label{supp:cost}

The paper measures a partitioning cost $c(K)$ that is increasing and convex in
$\log_2 K$, and attributes it to fill time and write locality. Two other
explanations are ruled out here.

\textbf{The cost does not come from write imbalance.} A natural explanation for
a partitioning cost is that a lopsided split starves one frontier: a class
holding 5\% of the address space but taking most of the writes should be forced
to reclaim blocks that are still full. We tested it directly, with
information-free labels so no benefit can offset the cost, sweeping the labelled
mass from 50\% to 2\% of the address space (uniform writes, 3 passes, two
seeds: the configuration of the pre-registered test T2, whose two balances it
reproduces exactly). Under $d{=}24$ the cost is indistinguishable from zero at
every balance: $+0.4\%$, $+0.5\%$, $+0.4\%$, $-0.0\%$, $+0.1\%$ at 50, 20, 10, 5
and 2\%. The thinnest class still spans about eleven reclaim
units, so this is not an artifact of a class too small to separate.

This matters because we measure a much larger cost for the same kind of
information-free split under a blind collector: $+11.1\%$ at 5\%, rising to
$+32.6\%$ at 20\%, and unmeasurable at 50\%, where FIFO cannot free a
frontier. Imbalance is ruled out as the explanation, which leaves the
collector and the rule by which capacity is shared between frontiers. The
capacity rule is tested below.
We have since implemented the equal-split rule, in which each frontier may hold
at most $\lceil N/(K{+}1)\rceil$ blocks and a frontier at its quota must reclaim
one of its own. \textbf{At this device size the rule does not bind}, so it costs
nothing:

\begin{center}
\begin{tabular}{rrrrr}
\toprule
$K$ & cap & max held & no rule & equal split \\
\midrule
2  & 200 & 76 & $+0.6\%$  & $+0.6\%$ \\
4  & 120 & 41 & $+1.8\%$  & $+1.8\%$ \\
8  & 67  & 23 & $+4.5\%$  & $+4.5\%$ \\
16 & 36  & 12 & $+10.4\%$ & $+10.4\%$ \\
\bottomrule
\end{tabular}
\end{center}

\noindent (Uniform writes, $S_f{=}0.10$, $d{=}24$, 12 passes, three seeds;
the no-rule column is the $c(K)$ of the paper.) No frontier ever holds more
than about a third of its quota --- 41 of 120 blocks at $K{=}4$ --- and the
capped and uncapped runs agree exactly. Under uniform
writes the rule only bites at far tighter quotas: a cap of 20 blocks costs
$+3.2\%$ and one of 10 blocks $+9.8\%$ at $d{=}24$ (three seeds). We do not
claim that tighter quotas always cost more. A frontier at its quota must
reclaim one of its own blocks, so a binding quota also changes the set the
collector chooses from --- the variable this paper is about --- and on the
skewed workload a binding quota lowers write amplification in some cells (by up
to $7.4\%$ at $N{=}600$ and $8.3\%$ at $N{=}200$, both under FIFO). Its cost is therefore reported for uniform writes
only.

This changes what we can say. A per-frontier quota at the natural setting is
\emph{not} the reason partitioning costs anything here --- the cost measured above is the fill-time and locality effect alone, not a capacity
charge. Whether a device that allocates more aggressively than
$\lceil N/(K{+}1)\rceil$, or holds a separate reserve per class, would impose
one is untested; the per-frontier-reserve rule grows linearly in $K$ and should.
The cost observed under blind collection tracks the collector. Holding the
labels, the balance and the device fixed and varying only victim selection, the
same information-free split costs $-0.0\%$ under $d{=}24$ and $+11.1\%$ under
FIFO. This comparison varies the collector alone; the capacity rule cannot
enter it, since HCWF has no per-frontier ownership for a rule to bind on. The
rule is tested where it can bind, in the $K$ family above, and does not bind at
its natural quota. With a balanced split under FIFO the collector cannot free a frontier
at this spare factor, and we report that cell as unmeasurable rather than omit
it. \textbf{The partitioning cost is itself a function of the collector}: a
selective collector routes around a poor split, a blind one pays for it.

\textbf{The capacity rule.} By default the simulator holds a single shared
free pool. That is not the ``proportional''
capacity rule --- it is the \emph{absence} of a capacity constraint: each
frontier takes whatever it needs. No device can implement this, since the device
cannot know future write ratios, and it is the rule most favourable to labelled
placement. Realisable rules --- an equal share of pools per frontier, or a
reserve held per frontier --- constrain frontiers and must cost at least as
much. \textbf{The cost figures below are therefore a lower bound}, and the
conclusion that the information term dominates holds under an idealisation
rather than under any rule a device offers. The equal-split rule is implemented and measured above; per-frontier reserve,
in which each class holds its own spare, is not, and would cost more again since
its overhead grows linearly in $K$.

\section{Class count across label constructions}
\label{supp:kstar}

We had expected $K^\ast$ itself to differ between baselines and it does not.
That prediction is not supported; what the baseline changes here is how much
the labels appear to be worth, not how many classes are profitable.

\textbf{$K^\ast$ is not robust across label constructions, and we report all
three.} On the headline workload with nested partitions the sweep gives
$+9.0\%, +8.3\%, +5.2\%, +1.0\%$ at $K{=}2,4,8,16$ against DWF --- monotone
decreasing, so $K^\ast{=}2$. Two other label constructions on related
workloads gave $K^\ast{=}4$ (nested partitions of a geometric workload, the
paper's class-count table) and, with rate-quantile labels, no class count worth
more than about 1\% ($+0.3\%$, $+0.3\%$, $+1.0\%$ at $K{=}2,4,8$). The three differ only in how the
classes are cut, which is the point: \textbf{the profitable class count is a
property of the labelling, not of the workload}, and a paper reporting a single
$K^\ast$ without stating its partition rule has not reported enough to be
reproduced. What all three share is that the benefit is largest at small $K$ and
falls away by $K{=}16$.

\textbf{On comparability with HCWF.} These families differ architecturally, not
just in partition. HCWF holds two frontiers and no separate collector frontier:
survivors go to the hot or cold frontier by the victim's label, and blocks are
relabelled on reclaim. $\mathrm{KWF}(K)$ holds $K$ data frontiers \emph{plus} a
dedicated collector frontier. KWF is therefore the DWF family extended to $K$
data streams, and is anchored by $\mathrm{KWF}(1) = \mathrm{DWF}$, which is
exact. It was never expected to reproduce HCWF, and the two-class figures
elsewhere in this paper are not comparable with this sweep.

\section{Simulator validation in full}
\label{supp:anchors}

\paragraph{Two analytic anchors}

The victim-selection axis is anchored at both ends by closed forms, neither
fitted to our data.

\textbf{Random selection.} If the collector chooses victims uniformly, the
expected valid fraction of the chosen block is the utilisation itself, so
$\mathrm{WAF} = 1/(1-u)$; at $u{=}0.90$ this is $10.00$. Measured, SWF on
uniform writes at $d{=}1$ (12 passes, three seeds): $9.984$, $0.2\%$ away.

\textbf{FIFO.} FIFO is not random selection. It takes the oldest block, which
has had a full queue cycle to accumulate invalidations, so its valid fraction
must fall \emph{below} the utilisation. In steady state a block is reclaimed after $N$
reclamations, each freeing $b(1-u)$ pages, giving the fixed point
\begin{equation}
u \;=\; \exp\!\Bigl(-\frac{1-u}{1-S_f}\Bigr),
\label{eq:fifo}
\end{equation}
which at $S_f{=}0.10$ yields $u{=}0.807$ and $\mathrm{WAF}{=}5.179$. Measured,
same configuration under FIFO: $5.241$, $1.2\%$ away. The ordering $u_{\text{FIFO}} < u_{\text{random}}$ is
itself a check the simulator passes. We note that
Eq.~\eqref{eq:fifo} is our derivation: the random-selection anchor reproduces a
known result, while the FIFO fixed point is validated against measurement and
against that ordering, not against an independent publication: we have not
located a published FIFO cleaning-cost derivation to compare against.

\paragraph{A label-independent consistency check}

Uniform random writes contain no rate structure, so no approach can separate
anything and all three should collapse to the same value. With meaningless
labels they do: SWF $5.241$, DWF $5.272$, HCWF $5.317$ --- a $1.01\times$ spread
around the closed form of $5.179$. This checks all three implementations at
once for any label-independent bias. Repeating with \emph{random} 5\% labels
raises HCWF to $6.423$, a $21\%$ penalty against its own meaningless-label
value: the partitioning cost of
Proposition 3 of the paper in isolation, paid where there is no information to
offset it.

\paragraph{Reproducing the two-class result}

The simulator reproduces the 2014 study before any new claim. On Rosenblum
$r{=}0.9$, $f{=}0.05$ at $d{=}10$: SWF 5.841, DWF 2.986, HCWF 2.187 --- the
published ordering. The DWF--HCWF gap narrows as the hot data gets hotter:
$+1.069$, $+0.797$, $+0.317$. Sweeping false positives and negatives
jointly, HCWF degrades 2.2049, 2.4462, 2.5825, 2.7167 and 3.0155 at 0--10\% and
crosses DWF (2.9912) between 6\% and 10\%, against a published threshold near
5\%~\cite{vanhoudt2014necessity}. Both frontier approaches show an interior optimum in $d$.

\paragraph{Dispersion of $g$ against the spare factor}

The paper withdraws a first-order bound on the dispersion of $g$ in favour of
measurement. The measured values ($r{=}0.9$, $f{=}0.05$, $d{=}24$, 12 passes,
five seeds) are:

\begin{center}
\begin{tabular}{lrrr}
\toprule
$S_f$ & $A_{\text{base}}$ & $g$ & s.d.($g$) \\
\midrule
0.05 & 5.76 & 0.489 & 0.0023 \\
0.10 & 3.39 & 0.506 & 0.0027 \\
0.20 & 2.01 & 0.560 & 0.0049 \\
0.30 & 1.51 & 0.625 & 0.0047 \\
0.40 & 1.25 & 0.695 & 0.0056 \\
\bottomrule
\end{tabular}
\end{center}

\paragraph{Run length on the robustness grid}

At the shorter run length of an earlier draft (4 passes), the same grid passed
in all 44 cells then reportable, with four cells suppressed by the
$A_{\text{base}}$ rule; we report the 12-pass result because 4 passes is
below the length at which the labelled scheme's amplification stops rising.

\section{Pre-registered verdict rules}
\label{supp:prereg}

Several results in the paper are described as pre-registered. The rules were
fixed in the test harness's docstrings before the corresponding runs and are
shipped with the artifact (\texttt{remaining.py}, \texttt{t9\_full.py},
\texttt{capeq.py}, \texttt{fifo\_robust.py}, \texttt{varmail\_robust.py}), so
they can be read against the commit that produced each table rather than taken
on trust. All six are reproduced here with their outcomes.

\textbf{Robustness grid (Sec.~IV-C).} The shape holds if the argmax of
$g$ is interior to the grid ($4<d^\ast<64$) and both the rise from $d{=}4$ and
the fall to greedy exceed $2\sigma$ (combined seed dispersion of the two
points), in every cell with $A_{\text{base}}\ge1.2$. The argmax was expected
to move between cells and its movement is not counter-evidence.
\emph{Outcome:} at 12 passes, passed in 47 of 48 cells; one cell failed with its
argmax at the grid's upper edge, tied with the next point. At 4 passes it passed
in all 44 cells then reportable. The separate expectation that $d^\ast$ shifts
with unit size failed.

\textbf{FIFO reversal (Sec.~IV-E).} Holds in a configuration if DWF
beats HCWF under FIFO in every workload seed. \emph{Outcome:} held in all six
configurations (two workloads, three device sizes).

\textbf{Capacity equivalent (Sec.~V-A and V-B).} Converged
when no entry moves by more than two points or two seed s.d.\ between the two
longest run lengths; the sign reversal holds if the value is positive at zero
error and negative at 20\%, each by more than two s.d. \emph{Outcome:}
converged between 24 and 32 passes; reversal held.

\textbf{Blind-collector convergence (Sec.~IV-E).} Converged if the last
step's change in mean is under 5\% and the seed dispersion is under 10\% of the
mean. \emph{Outcome:} with the three seeds the test was registered with, split
--- the change was $+5.5\%$ and the dispersion $1.9\%$ --- so the rule did not
decide it, and the classification was made on the shape of the sequence
instead. The paper's run-length table now uses five seeds, to match the other
FIFO results; with them the rule is met ($+3.1\%$, $1.9\%$), but because seeds
were added after the rule was fixed we do not count this as a pre-registered
pass.

\textbf{Two-class anchor.} $\mathrm{KWF}(1)$ must reproduce DWF exactly.
\emph{Outcome:} exact. A second anchor comparing $\mathrm{KWF}(2)$ with HCWF was
withdrawn as mis-specified: the two are different architectures, not two
partitions of one scheme.

\textbf{Trace acceptance (Sec.~III-F).} A real trace is admissible only
if its overwrite ratio is at least 3, $A_{\text{DWF}}\ge1.2$, and an oracle
label beats DWF. \emph{Outcome:} our RocksDB attempt failed the first and third
and is reported as a failure, not as a null result. On Varmail, the 2M- and
10M-write prefixes failed the first (1.99, 2.61) and were rejected; the
50M-write prefix passed all three (3.07, 1.704, 1.507 against 1.704) and is
reported in Sec.~IV-D. The two claims tested on it --- an interior
optimum in $g$, and a winner that changes with the collector --- were fixed
before the run; both held, with three seeds and again in the next 50M writes.

\section{Corrections to earlier drafts}
\label{supp:corrections}

This work went through several internal revisions in which measurements
overturned claims we had already written down. The corrections that change how a
result should be read are stated where the result appears. The remainder are
collected here so the body reads as the final account rather than as a
chronology.

\begin{itemize}
\item A monotonicity claim for conditional mutual information under refinement
of the conditioning $\sigma$-algebra was asserted with a Jensen-based proof. It
is false; see Sec.~III-D for the counterexample and the weaker statement
that replaces it.
\item Convexity of the partitioning cost was asserted in $K$ from plain second
differences on a grid that is evenly spaced only in $\log_2 K$. The cost is
clearly convex in $\log_2 K$; in $K$ convexity is marginal.
\item Several figures had no recorded run and were regenerated from stated
configurations (\texttt{regen\_unsourced.py}, \texttt{swf\_fifo.py}): the
partitioning-cost table (for example $0.4\%$, $1.7\%$, $4.8\%$, $10.4\%$ became
$0.6\%$, $1.8\%$, $4.5\%$, $10.4\%$), the quota table, the information-free
costs, the write-imbalance test, the rate-quantile class-count sweep and the
uniform-write anchors. The rate-quantile sweep had read $-0.2\%$, $-0.8\%$,
$+0.4\%$ and ``$K^\ast{=}1$''; it is $+0.3\%$, $+0.3\%$, $+1.0\%$, so rate labels
add at most about 1\%, not nothing.
\item SWF under FIFO was reported as not converging, swinging $55.6$, $99.8$
and $70.2$ across device sizes. The current simulator gives a stable $6.93$ at
all three sizes, and the earlier figures could not be reproduced; they are
withdrawn, along with the mechanism offered for them. The FIFO values of the
accounts table now come from one 32-pass, five-seed configuration, which moves
$g$ under FIFO at the headline device from $-1.53$ to $-1.72$.
\item Statistics of $\kd$ quoted from a grid whose raw output is not in the
archive (partial $R^2$ of $0.158$, $0.020$ and $0.086$; correlations of
$0.9908$, $0.9862$ and $-0.48$; a matched design of 28 groups) were replaced by
the values the archived sweeps give.
\item An analytic bound on the dispersion of $g$ predicted a $9.4\times$ growth
that measurement put at $2.1\times$, non-monotone. The bound is withdrawn in
favour of measured dispersion.
\item The ratio metric was said to rise monotonically to greedy at $r{=}0.95$
and so to contradict the interior optimum. That was read from a grid ending at
$d{=}16$; on the full grid the ratio peaks and falls as $g$ does
(Sec.~IV-B). A reversal of the DWF--HCWF ranking at $d{=}1$ and
$d{=}2$ was likewise withdrawn: over five seeds those rows are ties. The
reversal that stands is under FIFO (Sec.~IV-E).
\item The capacity equivalent was reported as $+68.6\%$ with perfect labels,
$-8.6\%$ at 20\% error and a break-even near 14\% error, from a run of about
six passes in which the labelled scheme had not reached steady state. At a
converged length the figures are $+83.2\%$, $-19.2\%$ and between 5\% and
10\%; the sign reversal was unaffected.
\item The capacity-equivalent table had an ``own-best'' column described as
running each scheme at its own best selectivity. It did not: it held HCWF at
the row's $d$ and moved only DWF to its best ($d{=}10$), so its $+54.6\%$ at
$d{=}24$ understated the labels. Running each scheme at its own best ($d{=}16$
for HCWF, $d{=}10$ for DWF) gives $+64.7\%$, which replaces it.
\item the collector sweep, the two-metric table and the numbers around them
had been typed from runs no script recorded; both tables are now produced by
scripts at one configuration, and the values changed by up to $0.05$ in $g$.
\item The FIFO run-length table used three workload seeds while the accounts
table and the robustness runs used five, so its 32-pass row ($18.159\pm1.337$)
disagreed with the $17.3\pm1.5$ quoted beside it. Both came from the same runs;
the table now uses the same five seeds, and its 32-pass row is $17.259\pm1.518$.
\item The membership experiment was described as bucketing cohorts into handles
by creation order; the code assigns cohort $c$ to handle $c \bmod K$, which
uses no death-order information. The text now says so. The experiment had also
run one seed at one collector; it now runs five seeds, $K$ from 1 to 32 and four
collectors (Table~\ref{supp:handles}). The sixteen-handle figures moved from
$+53.5\%$ and $+41.7\%$ (95\% and 77\% of the oracle) to $+52.9\%$ and
$+41.9\%$ (94\% and 75\%).
\item An empirical scaling $d^\ast \propto 1/f$ was refuted once the two
workload parameters were separated.
\item A co-death statistic reported in earlier work by the same authors was
withdrawn after it proved to measure the fraction invalidated rather than the
timing of invalidation.
\end{itemize}

\subsection{Two disclosures moved from the body}

\paragraph{How the argmax was resolved}

\textbf{The argmax is resolved in $g$, and was not in the ratio.} Successive
refinements of the $d$ grid moved the apparent peak from 16 to 24 to 32 in the
ratio metric, which we had read as grid-dependence to be reported as an
interval. In $g$, with dispersion measured over five seeds, the peak is a
short plateau with resolved edges: $0.506$ at $d{=}24$ and $0.505$ at $d{=}20$,
against $0.486$ at $d{=}16$ and $0.492$ at $d{=}32$, which lie $3.4\sigma$ and
$3.9\sigma$ below the peak, and $0.470$ at $d{=}14$ and $0.446$ at $d{=}48$,
more than $8\sigma$ below. Here $\sigma$ is the combined seed dispersion of
the two points compared.

This is a practical reason to prefer $g$ alongside the arithmetic one of
Sec.~III-B: in the ratio, $d{=}20$, $24$ and $32$ lie within $0.7$
percentage points of one another ($35.0\%$, $35.7\%$, $35.5\%$), so the ratio
blurs an argmax that $g$ resolves. The existence of a broad interior
optimum is robust; the location of the argmax is not, and any claim keyed to a specific $d$
should be read as keyed to that interval.

\paragraph{The figure we do not quote from the most recent system}

Every figure quoted here has been checked against the published text. One
number we do \emph{not} quote: the same section reports both a 25.1\% and a
15.5\% average reduction in one sentence, and only the latter is consistent with
the abstract and the corresponding figure, so we cite 15.5\% and 23.2\% and
leave the other alone. We have not re-run the system.
We cannot settle this from outside: re-running that
system at matched victim selection needs its code, not ours. We state the
conditional instead. If the reported gain survives at matched victim selection,
the placement attribution stands and our critique narrows to reporting practice
rather than to the result --- which is the outcome we would expect, and would
not weaken the paper's claim that the collector must be reported.

\section{Tabulated values behind the figures}
\label{supp:figdata}

Figures 2 and 3 of the paper are plotted from these values.

\begin{table}[t]
\caption{Capacity equivalent of a two-class labelling against realised label
error. $S_f{=}0.10$, $d{=}24$, three seeds, 32 passes.}
\label{supp:errcap}
\centering
\begin{tabular}{rrr}
\toprule
label error & $\Delta S/S$ & s.d. \\
\midrule
0.00 & $+83.2\%$ & 0.8 \\
0.05 & $+21.3\%$ & 3.1 \\
0.10 & $-2.2\%$  & 1.2 \\
0.20 & $-19.2\%$ & 0.5 \\
0.30 & $-25.8\%$ & 0.2 \\
0.44 & $-30.2\%$ & 0.2 \\
\bottomrule
\end{tabular}
\end{table}

\begin{table}[t]
\caption{Membership labels: write-amplification reduction over DWF against
handle count and collector, mean of five seeds. Top: 27 cohorts ($\rho{=}20$);
bottom: 54 cohorts ($\rho{=}10$).}
\label{supp:handles}
\centering
\small
\begin{tabular}{lrrrr}
\toprule
$K$ & $d{=}8$ & $d{=}24$ & greedy & FIFO \\
\midrule
\multicolumn{5}{l}{\emph{27 cohorts}}\\
1 & $0.0\%$ & $0.0\%$ & $0.0\%$ & $0.0\%$ \\
2 & $+6.7\%$ & $+9.3\%$ & $+6.3\%$ & $-0.8\%$ \\
4 & $+18.6\%$ & $+24.0\%$ & $+18.8\%$ & $-1.2\%$ \\
8 & $+28.6\%$ & $+40.9\%$ & $+35.5\%$ & $-3.2\%$ \\
16 & $+33.9\%$ & $+52.9\%$ & $+55.0\%$ & $-3.4\%$ \\
per cohort & $+34.4\%$ & $+56.4\%$ & $+63.2\%$ & $-12.6\%$ \\
\midrule
\multicolumn{5}{l}{\emph{54 cohorts}}\\
1 & $0.0\%$ & $0.0\%$ & $0.0\%$ & $0.0\%$ \\
2 & $+4.6\%$ & $+5.6\%$ & $+3.7\%$ & $-1.6\%$ \\
4 & $+11.4\%$ & $+14.5\%$ & $+9.7\%$ & $-2.7\%$ \\
8 & $+21.3\%$ & $+27.8\%$ & $+19.5\%$ & $-5.1\%$ \\
16 & $+30.2\%$ & $+41.9\%$ & $+34.6\%$ & $-10.8\%$ \\
32 & $+34.1\%$ & $+54.1\%$ & $+53.6\%$ & $-17.5\%$ \\
per cohort & $+26.8\%$ & $+55.8\%$ & $+65.8\%$ & $-41.0\%$ \\
\bottomrule
\end{tabular}
\end{table}

Table~\ref{supp:handles} is the full membership sweep behind Fig.~3
(\texttt{t8\_sweep.py}; $N{=}600$, $b{=}32$, $S_f{=}0.10$, 5 passes; seed s.d.\
at most $4.8$ points). Three points are not in the paper. At sixteen handles
the best collector depends on the workload: greedy is slightly ahead with 27
cohorts and $d{=}24$ with 54, because greedy needs nearly one handle per cohort
to reach its ceiling. At $d{=}8$ with 54 cohorts, one handle per cohort does
worse than 32 handles, because its 55 open blocks occupy almost all of the 60
free ones. And under FIFO every handle count costs, increasingly with $K$.

\section{The $\kd$ statistic in full}
\label{supp:further}

Proposition 1 of the paper predicts that a workload statistic measuring
within-pool invalidation structure cannot predict label value, because that
structure is $\mathcal{F}_C$-measurable. We built one and confirmed it.

$\kd$ is the largest fraction of a reclaim unit's blocks invalidated within a
window of $k$ fill durations, measured under a no-placement reference. Over 15
cohort workloads, its partial $R^2$ on the label-error crossover given the
zero-error gain is $0.151$ (\texttt{label\_tolerance.py --analyze}); in a
matched design that screened 180 candidates and kept 83 workloads in 26 groups
of equal zero-error gain, it explains $R^2 = 0.167$ of the crossover
(\texttt{matched\_pairs.py report}). The window does little: $\kd(k{=}1)$
and $\kd(k{=}50)$ correlate at $0.95$ on the concurrency sweep, $0.88$ on the
capacity-limited sweep and $0.86$ over the 180 screened candidates. And
$\operatorname{corr}(\kd, 1-u) = 0.97$ on the concurrency sweep (18 runs) and
$0.86$ on the capacity-limited sweep (27 runs), so $\kd$ largely tracks the invalidated fraction --- a permutation-invariant
count, exactly the quantity Proposition 1 of the paper says the collector
already has. Its correlation with available headroom is accordingly negative
($-0.58$ over the 27 capacity-limited runs).

\noindent\textbf{Recommendation.} Regress any statistic intended to capture
invalidation timing on $1-u$, and show it responds to its own window. The right
statistic measures \emph{cross-pool membership interleaving}, which the
collector cannot reach; cohorts-per-unit is a first attempt.

\end{document}